\documentclass[pra,showpacs]{revtex4}

\def\states{16\,000}

\usepackage{bm}
\usepackage{graphicx}
\usepackage{amssymb}

\begin{document}

\title{Entangling power of the quantum baker's map}

\author{A. J. Scott}
\email{ascott@info.phys.unm.edu} \affiliation{Department of Physics
and Astronomy, University of New Mexico, Albuquerque, NM 87131-1156,
USA}
\author{Carlton M. Caves}
\email{caves@info.phys.unm.edu} \affiliation{Department of Physics
and Astronomy, University of New Mexico, Albuquerque, NM 87131-1156,
USA}

\begin{abstract}
We investigate entanglement production in a class of quantum baker's
maps.  The dynamics of these maps is constructed using strings of
qubits, providing a natural tensor-product structure for application
of various entanglement measures.  We find that, in general, the
quantum baker's maps are good at generating entanglement, producing
multipartite entanglement amongst the qubits close to that expected
in random states.  We investigate the evolution of several
entanglement measures: the subsystem linear entropy, the concurrence
to characterize entanglement between pairs of qubits, and two
proposals for a measure of multipartite entanglement.  Also derived
are some new analytical formulae describing the levels of
entanglement expected in random pure states.
\end{abstract}

\pacs{05.45.Mt, 03.67.Mn}
\maketitle

\newcommand{\tr}{\,\text{tr}}
\def\ket#1{|#1\rangle}
\def\bra#1{\langle#1|}
\def\inner#1#2{\langle #1 | #2 \rangle}

\section{Introduction}
\label{sec1}

The introduction of ``toy'' mappings that demonstrate essential
features of nonlinear dynamics has led to many insights in the field
of classical chaos.  A well known example is the so-called {\it
baker's transformation\/} \cite{lichtenberg}.  It maps the unit
square, which can be thought of as a toroidal phase space, onto
itself in an area-preserving way.   Interest in the baker's map stems
from its straightforward characterization in terms of a Bernoulli
shift on the binary sequence that specifies a point in the unit
square.  It seems natural to consider a quantum version of the
baker's map for the investigation of quantum chaos. There is,
however, no unique procedure for quantizing a classical map; hence,
different quantum maps can lead to the same classical baker's
transformation.

Balazs and Voros \cite{balazs} were the first to formulate a quantum
version of the baker's map.  This was done with the help of the
discrete quantum Fourier transform.  Subsequently, improvements to
the Balazs-Voros quantization were made by Saraceno \cite{saraceno},
an optical analogy was found \cite{hannay}, a canonical quantization
was devised \cite{rubin,lesniewski}, and quantum computing
realizations were proposed \cite{schack2,brun}.  A related quantum
baker's mapping on the sphere has also been defined~\cite{pakonski}.

More recently, an entire class of quantum baker's maps, based on the
$2^N$-dimensional Hilbert space of $N$ qubits, was proposed by Schack
and one of us \cite{schack}.  The qubit structure provides a
connection to the binary representation of the classical baker's map.
This connection comes through the use of partial Fourier transforms,
which are used to define orthonormal basis states that are localized
on the unit-square phase space.  The $n$-th partial Fourier
transform, which acts on $N-n$ of the qubits, defines orthonormal
states that are localized at a lattice of phase-space points
specified by $n$ position bits and $N-n$ momentum bits.  Each state
is localized strictly within a position width $1/2^n$ and roughly
within a momentum width $1/2^{N-n}$.  The $n$-th quantum baker's map
in the class takes the states defined by the $n$-th partial Fourier
transform to the states defined by the $(n-1)$-th partial Fourier
transform.  This action decreases the number of position bits by one,
while increasing the number of momentum bits by one, thus mimicking
the stretching and squeezing of the classical baker's map. By using
this procedure, one obtains $N$ different quantum baker's maps, one
map for each number of initial position bits (or initial momentum
bits).  The Balazs-Voros quantization is but one member of this
class, corresponding to a single initial position bit ($n=1$).  The
map at the other extreme ($n=N$), which has no initial momentum bits,
is easily shown to be unentangling.

The classical limit of this class of baker's maps has been
investigated by two groups \cite{soklakov,tracy}.  Tracy and one of us
\cite{tracy} found that if the number of initial momentum bits is
allowed to approach infinity as the overall number of qubits goes to
infinity, the classical baker's map is recovered.  This result is
consistent with the findings of Soklakov and Schack \cite{soklakov}.
In contrast, if the number of momentum bits is held constant as the
number of qubits increases to infinity, a stochastic variation of the
classical baker's map is created \cite{tracy}.  The simplest such
limit, which holds the number of initial momentum bits constant at
zero ($n=N$), follows a sequence of completely unentangling quantum
baker's maps.

Our curiosity now poses the following two questions.  What is the
entangling power of {\it all\/} the quantum baker's maps?  And what
role, if any, does entanglement play in the the classical limit?
This paper focuses, for the most part, on the first of these
questions, investigating the entangling power of the Schack-Caves
class of quantum baker's maps.  Previous investigations of
entanglement in quantized chaotic systems, for the most part, have
dealt with the correlations induced by coupling two or more
independent systems together
\cite{sakagami,tanaka,furuya,angelo,miller,lakshminarayan,bandyopadhyay,%
tanaka2,fujisaki,lahiri}.  Our approach here is quite different: each
of our quantum baker's maps lives in a Hilbert space with a qubit
tensor-product structure; strings of qubits form a natural basis,
anchoring Hilbert space to the corresponding classical phase space,
and the quantum dynamics of our baker's maps is defined explicitly in
terms of this connection.  We therefore expect an intimate
relationship between the dynamics and the multipartite entanglement
induced amongst the qubits.  Using different approaches, the dynamics of
entanglement in qubit bases was recently investigated in
\cite{lakshminarayan2,bettelli}.

In order to calibrate the entanglement produced by the quantum
baker's map, we compare it with the entanglement of random pure
states drawn from the appropriate Hilbert space.  Thus a by-product
of our investigation is to derive some new exact formulae describing
the levels of entanglement expected in random pure states.  As
measures of entanglement, we examine in detail the subsystem linear
entropy, deriving formulae for the variance and third cumulant, and
two proposals for a multipartite entanglement measure, where formulae
for the mean and variance are given.  Pairwise (mixed-state)
entanglement between two qubits drawn from $N$ qubits is investigated
numerically, using the {\it concurrence\/} as the entanglement
measure.

The paper is organized as follows. In Sec.~\ref{sec2}, we introduce
the baker's map, both in classical and quantal form. Section
\ref{sec3} is devoted to discussing the measures of entanglement and
the entanglement of typical pure states.  In Sec.~\ref{sec4} we
explore the entangling power of the quantum baker's maps.  Finally,
in Sec.~\ref{sec5}, we provide a brief discussion of our results.

\section{The quantum baker's map}
\label{sec2}

The classical baker's map is a standard example of chaotic dynamics
\cite{lichtenberg}.  It is a symplectic map of the unit square onto
itself defined by
\begin{eqnarray}
q_{n+1} &=& 2 q_n - \lfloor 2q_n \rfloor\;, \label{classbake1}\\
p_{n+1} &=& \left(p_n + \lfloor2 q_n \rfloor \right)/2\;, \label{classbake2}
\end{eqnarray}
where $q,p \in [0,1)$, $\lfloor x \rfloor$ is the integer part of
$x$, and $n$ denotes the $n$-th iteration of the map. Geometrically,
the map stretches the unit square by a factor of two in the $q$
direction, squeezes by a factor of a half in the $p$ direction, and
then stacks the right half onto the left.

Interest in the baker's map is due mainly to the simplicity of its
{\it symbolic dynamics}. If each point of the unit square is
identified through its binary representation, $q = 0\!\cdot\!s_1 s_2
\ldots = \sum_{k=1}^\infty s_k 2^{-k}$ and $p= 0\!\cdot\! s_0 s_{-1}
\ldots = \sum_{k=0}^\infty s_{-k} 2^{-k-1}$ ($s_i\in\{0,1\}$), with a
bi-infinite symbolic string
\begin{equation}
s = \ldots s_{-2} s_{-1} s_0 \bullet s_1 s_2 s_3 \ldots \;,
\label{symbseq}
\end{equation}
then the action of the baker's map is to shift the position of the
dot by one point to the right,
\begin{equation}
s\rightarrow s' = \ldots s_{-2} s_{-1} s_0 s_1 \bullet s_2 s_3 \ldots \;.
\end{equation}

For a quantum-mechanical version of the map, we work in a
$D$-dimensional Hilbert space, ${\cal H}_D$, spanned by either the
position states $\ket{q_j}$, with eigenvalues $q_j=(j+\beta)/D$, or
the momentum states $\ket{p_k}$, with eigenvalues $p_k =(k+\alpha)/D$
($j,k=0,\ldots,D-1$). The constants $\alpha,\beta\in [0,1)$ determine
the periodicity of the space: $\ket{q_{j+D}} = e^{-2 \pi i
\alpha}\ket{q_j}$, $\ket{p_{k+D}} = e^{2\pi i\beta}\ket{p_k}$. Such
double periodicity identifies ${\cal H}_D$ with a toroidal phase
space.  Periodic boundary conditions correspond to $\alpha=\beta=0$,
and anti-periodic boundary conditions to $\alpha=\beta=1/2$; because
of other symmetry considerations, these are the only two cases of
interest.  The vectors of each basis are orthonormal,
$\inner{q_j}{q_k}=\inner{p_j}{p_k}=\delta_{jk}$, and the two bases
are related {\it via\/} the finite Fourier transform,
\begin{equation}
\bra{q_j}\hat{F}_D\ket{q_k}\equiv\inner{q_j}{p_k}=
\frac{1}{\sqrt{D}}\,e^{iq_jp_k/\hbar}\;.
\end{equation}
For consistency of units, we must have $2 \pi \hbar D=1$.

The first work on a quantum baker's map was done by Balazs and Voros
\cite{balazs}.  Assuming an even-dimensional Hilbert space with
periodic boundary conditions, they defined a quantum baker's map in
terms of the unitary operator $\hat B$ that executes a single
iteration of the map.  To define the Balazs-Voros unitary operator in
our notation, imagine that the even-dimensional Hilbert space is a
tensor product of a qubit space and the space of a ($D/2$)-dimensional
system. Writing $j=x(D/2)+j'$, $x\in\{0,1\}$, we can write the
position eigenstates as $\ket{q_j}=\ket{x}\otimes\ket{j'}$, where the
states $\ket{x}$ make up the standard basis for the qubit, and the
states $\ket{j'}$ are a basis for the ($D/2$)-dimensional system.  The
state of the qubit thus determines whether the position eigenstate
lies in the left or right half of the unit square.  The Balazs-Voros
quantum baker's map is defined by
\begin{equation}
\hat{B}=\hat{F}_D\circ\Bigl(\hat 1_2\otimes\hat F_{D/2}^{-1}\Bigr)\;,
\label{BV}
\end{equation}
where $\hat 1_2$ is the unit operator for the qubit, and
$\hat{F}_{D/2}$ is the finite Fourier transform on the
($D/2$)-dimensional system.  The unitary $\hat B$ does separate
inverse Fourier transforms on the left and right halves of the unit
square, followed by a full Fourier transform.  Later Saraceno
\cite{saraceno} improved certain symmetry characteristics of this
quantum baker's map by using anti-periodic boundary conditions.

Taking again the anti-periodic Hilbert space (which we use throughout
the remainder of this paper), Schack and one of us \cite{schack}
introduced a class of quantum baker's maps for dimensions $D=2^N$.
For these cases, we can model our Hilbert space as the the space of
$N$ qubits, and the position states can be defined as product states
for the qubits in the standard basis, i.e.,
\begin{equation}
\ket{q_j}=
\ket{x_1}\otimes\ket{x_2}\otimes\cdots\otimes\ket{x_N}\;,
\label{qj}
\end{equation}
where $j$ has the binary expansion
\begin{equation}
j=x_1\ldots x_N\!\cdot\! 0=\sum_{l=1}^N x_l 2^{N-l}
\end{equation}
and $q_j=(j+1/2)/D=0\!\cdot\! x_1\ldots x_N1$.

The connection with the classical baker's map derives from the
symbolic dynamics.  The bi-infinite strings~(\ref{symbseq}) that
specify points in the unit square are replaced by sets of orthogonal
quantum states created through the use of a partial Fourier transform
\begin{equation}
\hat G_n\equiv\hat 1_{2^n}\otimes\hat F_{2^{N-n}}\;,\qquad n=0,\ldots,N,
\end{equation}
where $\hat 1_{2^n}$ is the unit operator on the first $n$ qubits and
$\hat F_{2^{N-n}}$ is the Fourier transform on the remaining qubits.
The partial Fourier transform thus transforms the $N-n$ least
significant qubits of a position state,
\begin{equation}
\label{partialfourier}
\hat{G}_n\,
\ket{x_1}\otimes\cdots\otimes\ket{x_n}\otimes
\ket{a_1}\otimes\cdots\otimes\ket{a_{N-n}}
=\,\ket{x_1}\otimes\cdots\otimes\ket{x_n}\otimes
\frac{1}{\sqrt{2^{N-n}}}\sum_{x_{n+1},\ldots,x_N}\ket{x_{n+1}}
\otimes\cdots\otimes\ket{x_N}e^{2\pi iax/2^{N-n}}\;,
\end{equation}
where $a$ and $x$ are defined through the binary representations
$a=a_1\ldots a_{N-n}\!\cdot\! 1$ and $x=x_{n+1}\ldots x_N\!\cdot\! 1$. In the
limiting cases, we have $\hat{G}_0=\hat{F}_{D}$ and
$\hat{G}_N=i\hat{1}$.  The analogy to the classical case is made
clear by introducing the following notation for the partially
transformed states:
\begin{equation}
\ket{\,a_{N-n}\ldots a_1\bullet x_1\ldots x_n}
\,\equiv\,\hat{G}_n\,\,\ket{x_1}\otimes\cdots\otimes\ket{x_n}\otimes
\ket{a_1}\otimes\cdots\otimes\ket{a_{N-n}}\;.
\label{partialfourierstates}
\end{equation}
For each value of $n$, these states form an orthonormal basis and
are localized in both position and momentum.  The state
$\ket{a_{N-n}\ldots a_1\bullet x_1\ldots x_n}$ is strictly localized
in a position region of width $1/2^n$ centered at $0\!\cdot\! x_1 \ldots
x_n 1$ and is roughly localized in a momentum region of width
$1/2^{N-n}$ centered at $0\!\cdot\! a_1 \ldots a_{N-n} 1$.  In the
notation of Eq.~(\ref{symbseq}), it is localized at the phase-space
point $1a_{N-n} \ldots a_1 \bullet x_1 \ldots x_n 1$.  Notice that
$\ket{a_{N}\ldots a_1\bullet}=
\hat{G}_0\,\ket{a_1}\otimes\cdots\otimes\ket{a_{N}}$ is a momentum
eigenstate and that $\ket{\!\bullet x_1\ldots x_N}=
\hat{G}_N\,\ket{x_1}\otimes\cdots\otimes\ket{x_N}=\,
i\ket{x_1}\otimes\cdots\otimes\ket{x_N}$ is a position eigenstate,
the $i$ being a consequence of the anti-periodic boundary conditions.

\begin{figure}[t]
\includegraphics[scale=1]{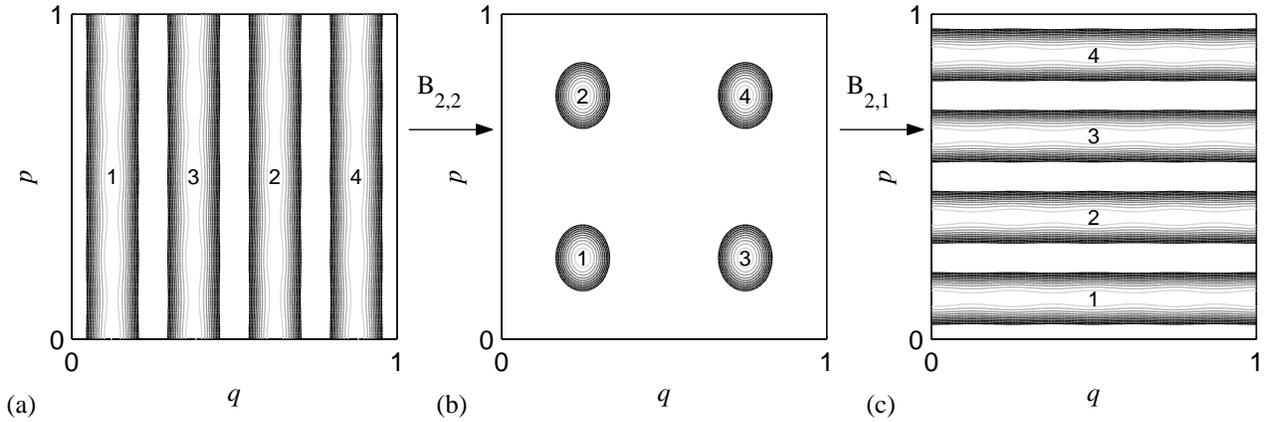}
\caption{Husimi function for each partially Fourier transformed
state (\ref{partialfourierstates}) when $N=2$: (a) $n=2$, (b) $n=1$,
and (c) $n=0$.  The action of the quantum baker's map $\hat B_{2,2}$
is to map the four states in (a) to the four states in (b), as shown
by the numbers labeling the states.  Similarly, the action of $\hat
B_{2,1}$ is to map the states in (b) to the states in (c).}
\label{fig1}
\end{figure}

Using this notation, a quantum baker's map on $N$ qubits is defined
for each value of $n=1,\ldots,N$ by the single-iteration unitary
operator \cite{schack}
\begin{equation}
\label{bakern}
\hat{B}_{N,n}\,\equiv\,
\hat{G}^{\phantom{-1}}_{n-1} \circ \hat{S}_n \circ \hat{G}^{\,-1}_n \; =
\sum_{x_1,\dots,x_n}\sum_{a_1,\dots,a_{N-n}}
\ket{\,a_{N-n}\dots a_1 x_1\bullet x_2\dots x_n} \bra{a_{N-n}\dots a_1
\bullet x_1 x_2\dots x_n}\;,
\end{equation}
where the shift operator $\hat{S}_n$ acts only on the first $n$
qubits, i.e., $\hat{S}_n\ket{x_1}\otimes\ket{x_2}\otimes\cdots\otimes
\ket{x_n}\otimes\ket{x_{n+1}}\otimes\cdots\otimes\ket{x_N}
=\ket{x_2}\otimes\cdots\otimes\ket{x_n}\otimes\ket{x_1}
\otimes\ket{x_{n+1}}\otimes\cdots\otimes\ket{x_N}$.  Notice that
since $\hat S_n$ commutes with $\hat G_n^{\,-1}$, we can put $\hat
B_{N,n}$ in the form
\begin{equation}
\hat B_{N,n}=\hat 1_{2^{n-1}}\otimes
\Bigl(\hat F_{2^{N-n+1}}\circ(\hat 1_2\otimes\hat F_{2^{N-n}}^{-1})\Bigr)
\circ\hat S_n\;.
\end{equation}
Since $\hat S_1$ is the unit operator, it is clear that $\hat B_{N,1}$
is the Balazs-Voros-Saraceno quantum baker's map~(\ref{BV}).

We can also write
\begin{equation}
\hat B_{N,n}=\hat 1_{2^{n-1}}\otimes\hat B_{N-n+1,1}\circ\hat S_n\;,
\end{equation}
which shows that the action of $\hat B_{N,n}$ is a shift of the $n$
leftmost qubits followed by application of the Balazs-Voros-Saraceno
baker's map to the $N-n+1$ rightmost qubits.  At each iteration, the
shift map $\hat S_n$ does two things: it shifts the $n$-th qubit, the
most significant qubit that was subject to the previous application
of $\hat B_{N-n+1,1}$, out of the region subject to the next
application of $\hat B_{N-n+1,1}$, and it shifts the most significant
(first) qubit into the region of subsequent application of $\hat
B_{N-n+1,1}$.

The quantum baker's map $\hat B_{N,n}$ takes a state localized at
$1a_{N-n} \ldots a_1 \bullet x_1 \ldots x_n 1$ to a state localized
at $1a_{N-n} \ldots a_1 x_1 \bullet x_2 \ldots x_n 1$.  The decrease
in the number of position bits and increase in momentum bits enforces
a stretching and squeezing of phase space in a manner resembling the
classical baker's map.  In Fig.~\ref{fig1}(a), (b), and (c), we plot
the Husimi function (defined as in \cite{tracy}) for the partially
Fourier transformed states (\ref{partialfourierstates}) when $N=2$,
and $n=2$, 1 and 0, respectively.  The quantum baker's map is a
one-to-one mapping of one basis to another, as shown in the
figure.

One useful representation of our quantum baker's maps, introduced in
\cite{schack}, starts from using standard techniques
\cite{nielsenchuang} to write the partially transformed
states~(\ref{partialfourier}) as product states:
\begin{equation}
\ket{\,a_{N-n}\ldots a_1\bullet x_1\ldots x_n}\,=\,
e^{\pi i(0\!\cdot\! a_1\ldots a_{N-n}1)}
\Biggl(\,\bigotimes_{k=1}^n\ket{x_k}\Biggr)
\Biggl(\,\bigotimes_{k=n+1}^N{1\over\sqrt2}\Bigl(
\ket{0}+e^{2\pi i(0\!\cdot\! a_{N-k+1}\ldots a_{N-n}1)}\ket{1}\Bigr)
\Biggr)\;.
\end{equation}
These input states are mapped by $\hat B_{N,n}$ to output states
\begin{eqnarray}
\ket{\,a_{N-n}\ldots a_1x_1\bullet x_2\ldots x_n}\,=\,&\mbox{}&
e^{\pi i(0\!\cdot\! x_1a_1\ldots a_{N-n}1)}
\Biggl(\,\bigotimes_{k=2}^n\ket{x_k}\Biggr) \nonumber \\
&\mbox{}&\quad\Biggl(\,\bigotimes_{k=n+1}^N
{1\over\sqrt2}\Bigl(
\ket{0}+e^{2\pi i(0\!\cdot\! a_{N-k+1}\ldots a_{N-n}1)}\ket{1}\Bigr)
\Biggr)
\otimes
{1\over\sqrt2}\Bigl(
\ket{0}+e^{2\pi i(0\!\cdot\! x_1a_1\ldots a_{N-n}1)}\ket{1}\Bigr)
\;. \nonumber \\
\end{eqnarray}
These forms show that the quantum baker's map $\hat B_{N,n}$ shifts the
states of all the qubits to the left, except the state of the
leftmost, most significant qubit.  The state $\ket{x_1}$ of the
leftmost qubit can be thought of as being shifted to the rightmost
qubit, where it suffers a controlled phase change that is determined
by the state parameters $a_1,\ldots,a_{N-n}$ of the original
``momentum qubits.''  The quantum baker's map can thus be written as
a shift map on a finite string of qubits, followed by a controlled
phase change on the least significant qubit.  Soklakov and Schack
\cite{soklakov} have developed this shift representation into a
useful tool.  Using an approach based on coarse graining in this
representation, they investigated the classical limit of the quantum
baker's maps.

Another useful representation of our quantum baker's maps is the
qubit (position) representation
\begin{eqnarray}
\hat{B}_{N,n}=\frac{\sqrt{2}}{2^{N-n+1}}
\!\!\!\!\sum_{\begin{array}{c}
\scriptstyle x_1,\dots,x_n \\
\scriptstyle a_1,\dots,a_{N-n}
\end{array}}
\!\!\sum_{\begin{array}{c}
\scriptstyle y_1,\dots,y_{N-n} \\
\scriptstyle z_1,\dots,z_{N-n+1}
\end{array}}
\!\!\!\!&\mbox{}&\ket{x_2}\bra{x_1}\otimes\cdots\otimes\ket{x_n}\bra{x_{n-1}}
\otimes
\ket{z_1}\bra{x_n}\otimes\ket{z_2}
\bra{y_1}\otimes\cdots\otimes\ket{z_{N-n+1}}\bra{y_{N-n}} \nonumber\\
&&\times
\exp\!\left[
\frac{\pi i}{2^{N-n}}\Big((j+1/2)(l+1/2)+
2^{N-n}x_1(l+1/2)-2(j+1/2)(k+1/2)\Big)
\right]\;,\nonumber \\
\end{eqnarray}
where
\[
j = \sum_{k=1}^{N-n}a_k 2^{N-n-k} \text{,}\quad
k = \sum_{k=1}^{N-n}y_k 2^{N-n-k} \text{,}\quad\text{and}\quad
l = \sum_{k=1}^{N-n+1}z_k 2^{N-n+1-k}.
\]
Soklakov and Schack \cite{soklakov} have found simplified forms
of this qubit representation, suitable for asymptotic analysis of
the classical limit.

The classical limit for the above quantum baker's maps has also been
investigated in \cite{tracy}, using an analysis based on the limiting
behavior of the coherent-state propagator of $\hat B_{N,n}$.   When
$D=2^N\rightarrow\infty$, the total number of qubits $N$ necessarily
becomes infinite, but one has a considerable choice in how to take
this limit.  For example, we could use only one position bit, thus
fixing $n=1$, and let the number of momentum bits $N-1$ become large.
This is the limiting case of the Balazs-Voros-Saraceno quantization.
There is, however, a wide variety of other scenarios to consider,
e.g., $n=N/2$ or $n=2N/3-1$ as $N\rightarrow\infty$.  In \cite{tracy}
it was shown that provided the number of momentum bits $N-n$
approaches infinity, the correct classical behavior is recovered in
the limit.  If the number of momentum bits remains constant, e.g.,
$n=N$ as $N\rightarrow\infty$, a stochastic variant of the classical
baker's map is found.  These results are consistent with those
obtained previously by Soklakov and Schack \cite{soklakov}.

The special case $n=N$, which does not limit to the classical baker's
map, has other interesting properties.  Although all
finite-dimensional unitary operators are quasi-periodic, the quantum
baker's map $\hat{B}_{N,N}$ is strictly periodic,
\begin{equation}
\hat{B}_{N,N}^{\,4N}=1\;, \label{B4N}
\end{equation}
as we show below.  All its eigenvalues, therefore, are $4N$-th roots
of unity, i.e., of the form $e^{\pi ik/2N}$, and, hence, there are
degeneracies when $N>4$.  This represents a strong deviation from the
predictions of {\it random matrix theory\/} \cite{haake,guhr,brody}.
The property (\ref{B4N}) can be easily shown by noting that $\hat
B_{N,N}=-i\hat G_{N-1}\circ\hat S_N= -i(\hat 1_{2^{N-1}}\otimes\hat
F_2)\circ\hat S_N$; i.e., $\hat B_{N,N}$ is a shift followed by
application of the unitary
\begin{equation}
\hat{U}\equiv-i\hat F_2=
{1\over\sqrt2}\Bigl(
e^{-\pi i/4}(\ket{0}\bra{0}+\ket{1}\bra{1})+
e^{\pi i/4}(\ket{0}\bra{1}+\ket{1}\bra{0})
\Big)
=e^{-\pi i/4}e^{i\hat{\sigma}_x(\pi/4)}
\end{equation}
to the least significant qubit.  On product states, the action of
$\hat B_{N,N}$ can be written explicitly as
\begin{equation}
\hat B_{N,N}\ket{\psi_1}\otimes\ket{\psi_2}\otimes\dots\otimes\ket{\psi_N}=
\ket{\psi_2}\otimes\dots\otimes\ket{\psi_N}\otimes\hat{U}\ket{\psi_1}\;.
\label{BN}
\end{equation}
Since $\hat U_1^4=1$, we get the property~(\ref{B4N}).  One can now
also see that $\hat{B}_{N,N}$ cannot entangle initial product states.

The eigenstates of $\hat U$ are $(\ket{0}+\ket{1})/\sqrt2$, with
eigenvalue $+1$, and $(\ket{0}-\ket{1})/\sqrt2$, with eigenvalue
$-i$.  For the discussion in this paragraph, label these eigenstates
by their eigenvalues $\alpha=1,-i$. We can construct eigenstates of
$\hat B_{N,N}$ from tensor products of these eigenstates.   Let $P$
denote the period of a string $\alpha_1\ldots\alpha_N$ under cycling;
for the the corresponding product state
$\ket{\psi}=\ket{\alpha_1}\otimes\cdots\otimes\ket{\alpha_N}$, $P$ is
the smallest positive integer such that $\hat
S_N^{\,P}\ket{\psi}=\ket{\psi}$.  It is now easy to show that the
eigenstates of $\hat B_{N,N}$ have the form
\begin{equation}
\ket{\Psi}=
{1\over\sqrt P}\sum_{k=0}^{P-1}
[\alpha_1\cdots\alpha_P]_P^{-k}\alpha_1\cdots\alpha_k\hat S_N^k
\ket{\alpha_1}\otimes\cdots\otimes\ket{\alpha_N}\;,
\label{eig}
\end{equation}
where $[\alpha]_P$ denotes a $P$-th root of $\alpha$.  The eigenvalue
of the state~(\ref{eig}) is $[\alpha_1\cdots\alpha_P]_P$.  Notice
that the product state $\ket{\alpha}^{\otimes N}$ ($\alpha=1,-i$) is
always an eigenstate, with eigenvalue $\alpha$.  As an example, the
eigenstates of $\hat B_{2,2}$ are
\begin{equation}
\ket{1}\otimes\ket{1}\;,
\quad
\ket{-i}\otimes\ket{-i}\;,
\quad
{1\over\sqrt2}
\Bigl(\ket{1}\otimes\ket{-i}+e^{\pi i/4}\ket{-i}\otimes\ket{1}\Bigr)\;,
\quad
{1\over\sqrt2}
\Bigl(\ket{1}\otimes\ket{-i}+e^{-3\pi i/4}\ket{-i}\otimes\ket{1}\Bigr)\;,
\end{equation}
with eigenvalues $1$, $-i$, $e^{-\pi i/4}$, and $e^{3\pi i/4}$,
respectively.

When $n<N$, the action of the quantum baker's map is similar to
(\ref{BN}), but with a crucial difference.  After the qubit string is
cycled, instead of applying a unitary to the rightmost, least
significant qubit, a joint unitary is applied to all of the the
$N-n+1$ rightmost qubits.  As discussed above, this joint unitary can
be realized as controlled phase change of the least significant
qubit, where the control is by the state parameters
$a_1,\ldots,a_{N-n}$ of the original momentum qubits.  This
controlled phase change means that initial product states become
entangled.  The resulting entanglement production is the subject of
this paper and is investigated in Sec.~\ref{sec4}.  Since the
entangling controlled-phase change involves an increasing number of
qubits as $n$ decreases from $n=N$ to $n=1$ (the
Balazs-Voros-Saraceno map), we might expect the entanglement to
increase as $n$ ranges from $N$ to 1.  What we find, however, is that
all the maps for $n$ not too close to $N$ are efficient entanglement
generators, but that the greatest entanglement is produced when $n$
is roughly midway between $N$ and 1.

To calibrate our entanglement production results, in the next section
we establish how much entanglement to expect for pure states chosen
randomly from the Hilbert space.  With the entanglement of typical
states quantified, we have a standard against which to compare the
entanglement produced by the quantum baker's map.  We might expect
the quantum baker's maps to be good at creating typical states in
Hilbert space, and our work on entanglement production can also be
regarded as a way of investigating this expectation.

\section{Entanglement of typical pure states}
\label{sec3}

Quantifying the amount of entanglement between quantum systems is a
recent pursuit that has attracted a diverse range of researchers
\cite{horodecki,wootters,horodecki2,nielsen}. The best understood
case, not surprisingly, is the simplest. It is generally accepted
that when a bipartite quantum system is in an overall pure state,
there is an essentially unique resource-based measure of entanglement
between the two subsystems.  This measure is given by the von Neumann
entropy of the marginal density operators \cite{bennett,popescu}.
Thus an investigation into typical values expected for the {\it
entropy of entanglement\/} seems a worthwhile endeavor in its own
right \cite{bandyopadhyay,zyczkowski}, which we undertake in
Sec.~\ref{sec3a}.  Notice that given an $N$-qubit quantum-baked
state, there are $2^{N-1}-1$ different possible partitions into the
two subsystems, and hence, $2^{N-1}-1$ different entropies of
entanglement to consider.

Another well understood case is the pairwise entanglement of two
qubits in an overall mixed state.  When a bipartite quantum system is
in a mixed state, proposals for measuring the entanglement include
the {\it entanglement of formation\/} \cite{wootters,bennett}, {\it
distillable entanglement\/} \cite{bennett,bennett2}, and {\it
relative entropy\/} \cite{vedral,vedral2}.  For pure states all these
reduce to the von Neumann entropy, but the first is the best
understood for mixed states.  In the case of two qubits in a mixed
state, an exact expression for the entanglement of formation exists
in terms of another measure called the {\it concurrence\/}
\cite{wootters,hill,wootters2}.  Thus another entanglement measure to
consider could be the concurrence that results after all but two
qubits are traced out of an $N$-qubit quantum-baked state.
Section~\ref{sec3b} is concerned with this pairwise entanglement.

Unlike the above special cases, quantifying the amount of
multipartite entanglement in a general multipartite state remains far
from being completely understood.  There have, nevertheless, been a
number of proposals for such a measure.  In Section \ref{sec3c} we
investigate two of these.  We must stress, however, that no single
measure alone is enough to quantify the entanglement in a
multipartite system.  As the number of subsystems increases, so too
does the number of independent entanglement measures.

\subsection{Bipartite pure-state entanglement}
\label{sec3a}

Consider a bipartite quantum system with Hilbert space
$\mathcal{H}_A\otimes\mathcal{H}_B$ of dimension $\mu\nu$, where
$\mu$ and $\nu$ are the dimensions of subsystems $A$ and $B$, with
$\mu\leq \nu$.  A joint pure state $\hat\rho=|\psi\rangle\langle\psi|$
has a Schmidt decomposition
$|\psi\rangle=\sum_{i=1}^\mu\sqrt{p_i}\,|a_i\rangle\otimes|b_i\rangle$,
where $|a_i\rangle$ and $|b_i\rangle$ are orthonormal bases spanning
$\mathcal{H}_A$ and $\mathcal{H}_B$.  If we sample random pure states
according to the unitarily invariant Haar measure, then the Schmidt
coefficients $0< p_i\leq 1$ obey the distribution \cite{lloyd}
\begin{equation}
P\left(p_1,\dots,p_\mu\right)dp_1\dots dp_\mu
= N\,\delta\!\!\left(1-\sum_{l=1}^\mu p_{\,l}\right)
\prod_{1\leq i<j\leq \mu}\left(p_i-p_j\right)^2
\prod_{k=1}^\mu p_k^{\nu-\mu}dp_k\;,
\label{distr}
\end{equation}
where $N$ is a normalization constant.

Considered as eigenvalues of the marginal density matrices
$\hat\rho_A=\tr_B\,\hat\rho$ and $\hat\rho_B=\tr_A\,\hat\rho$, the
Schmidt coefficients give the {\it von Neumann entropy\/} $S$ of each
subsystem
\begin{equation}
S=S_A=S_B\equiv-\tr\,\hat\rho_B\ln\hat\rho_B=-\sum_{i=1}^\mu p_i\ln p_i.
\end{equation}
As remarked above, the von Neumann entropy is generally considered to
be the unique resource-based measure of entanglement for a bipartite
quantum system in an overall pure state.  Given the distribution
(\ref{distr}), an expression for the average entropy can be
calculated:
\begin{equation}
S_{\mu,\nu}\equiv\langle S\rangle=
\sum_{k=\nu+1}^{\mu\nu}\frac{1}{k}-\frac{\mu-1}{2\nu}\;.
\end{equation}
This succinct formula was conjectured by Page \cite{page} and later
proved by others \cite{foong,ruiz,sen} (see also
\cite{bandyopadhyay,zyczkowski,gemmer,malacarne}).

An expression for the average {\it purity},
\begin{equation}
R=R_A=R_B\equiv\tr\,\hat\rho_B^2=\sum_{i=1}^\mu p_i^{\,2}\;,
\end{equation}
had been calculated much earlier by Lubkin \cite{lubkin}:
\begin{equation}
R_{\mu,\nu}\equiv\langle R\rangle=\frac{\mu+\nu}{\mu\nu+1}\;. \label{lubkinformula}
\end{equation}
The purity provides the first nontrivial term in a Taylor series
expansion of the von Neumann entropy about its maximum and because of
its simplicity, is much easier to investigate analytically.  For
these reasons we restrict our attention to it.

One can also define a {\it linear entropy\/} in terms of the purity:
\begin{equation}
S_L\equiv\beta(1-R)\;.
\end{equation}
We choose the normalization factor so that $0\leq S_L\leq 1$, i.e.,
$\beta\equiv \mu/(\mu-1)$. The average linear entropy is
\begin{equation}
\langle S_L\rangle=1-{\mu+1\over\mu\nu+1}\;,
\label{meanlarge}
\end{equation}
which shows that for any division into two subsystems, when the overall
dimension $\mu\nu$ is large, a typical state has nearly maximal
entanglement.

Ideally, we would like an expression for the complete probability
distribution $P(R)\,dR$ of the purity for random pure states.  This
function cannot, in general, be calculated analytically, so we are
forced to settle for formulae describing a few of the cumulants. For
subsystems of even moderately high dimension, the cumulants we
calculate are sufficient to describe accurately the entire
distribution $P(R)$.  In deriving these cumulants, we follow the work
of Sen \cite{sen}.

Consider the second moment
\begin{equation}
\langle R^2\rangle=
\int\sum_{i,j=1}^\mu p_i^{\,2}p_j^{\,2}P(\mathbf{p})\,d\mathbf{p}\;,
\label{R2}
\end{equation}
where $\mathbf{p}\equiv(p_1,\dots,p_\mu)$ and $d\mathbf{p}\equiv
dp_1\dots dp_\mu$.  We first remove the obstacle of integrating over
the probability simplex by noting that
\begin{equation}
Q(\mathbf{q})\,d\mathbf{q}
\equiv \prod_{1\leq i<j\leq \mu}\left(q_i-q_j\right)^2
\prod_{k=1}^\mu e^{-q_k}q_k^{\nu-\mu}\,dq_k \label{Q}\\
=N\,e^{-r}r^{\mu\nu-1}P(\mathbf{p})\,d\mathbf{p}\,dr\;,
\end{equation}
where the new variables $q_i\equiv rp_i$ take on values independently
in the range $[0,\infty)$.  Integrating over all the values of the
new variables, we find that the normalization constant is given by
$N=\overline Q/\Gamma(\mu\nu)$, where $\overline{Q}\equiv\int
Q(\mathbf{q})d\mathbf{q}$.  Similarly, we find that
\begin{equation}
\int q_i^2q_j^2Q(\mathbf{q})d\mathbf{q} =
\overline Q\,
\frac{\Gamma(\mu\nu+4)}{\Gamma(\mu\nu)}
\int p_i^{\,2}p_j^{\,2}P(\mathbf{p})\,d\mathbf{p}\;.
\label{QtoP}
\end{equation}
thus determining the desired second moment in terms of integrals
over $Q(\bf q)$.

Now notice that the first product in Eq.~(\ref{Q}) is the square of the
Van der Monde determinant
\begin{equation}
\Delta(\mathbf{q}) \,\equiv\,
\prod_{1\leq i<j\leq \mu}\left(q_i-q_j\right)
= \left| \begin{array}{ccc}
 1 & \ldots & 1 \\
 q_1 & \ldots & q_\mu \\
 \vdots & \ddots & \vdots \\
 q_1^{\mu-1} & \ldots & q_\mu^{\mu-1}
 \end{array} \right|
 = \left| \begin{array}{ccc}
 r_0^{\alpha}(q_1) & \ldots & r_0^{\alpha}(q_\mu) \\
 r_1^{\alpha}(q_1) & \ldots & r_1^{\alpha}(q_\mu) \\
 \vdots & \ddots & \vdots \\
 r_{\mu-1}^{\alpha}(q_1)  & \ldots & r_{\mu-1}^{\alpha}(q_\mu)
 \end{array} \right|\;.
 \label{Van2}
\end{equation}
The second determinant in Eq.~(\ref{Van2}) follows from the basic property of
invariance after adding a multiple of one row to another, with
$\alpha\equiv\nu-\mu$ and the polynomials
$r^\alpha_k(q)\equiv k!L^\alpha_k(q)$ judiciously chosen to be rescaled Laguerre
polynomials \cite{gradshteyn}, satisfying the recursion relation
\begin{equation}
r^\alpha_k(q) = r^{\alpha+1}_k(q)-kr^{\alpha+1}_{k-1}(q)
= \sum_{i=0}^j(-1)^i{\,j\,\choose i}
k(k-1)\dots(k-i+1) r^{\alpha+j}_{k-i}(q)
\label{L3}
\end{equation}
and having the orthogonality property
\begin{equation}
\int_0^\infty dq\,e^{-q}q^\alpha r^\alpha_k(q)r^\alpha_l(q) =
\Gamma(k+1)\Gamma(\alpha+k+1)\delta_{kl}\;.
\label{L1}
\end{equation}

These facts in hand, we can evaluate
\begin{eqnarray}
\overline{Q} &=&
\int\Delta(\mathbf{q})^2\prod_{k=1}^\mu e^{-q_k}q_k^\alpha\,dq_k \nonumber\\
&=&\mathop{\sum_{i_1,\dots,i_\mu}}_{j_1,\dots,j_\mu}
\epsilon_{i_1\dots i_\mu}\epsilon_{j_1\dots j_\mu}
\prod_{k=1}^\mu\int dq_k\,e^{-q_k}q_k^\alpha
r^\alpha_{i_k-1}(q_{k})r^\alpha_{j_k-1}(q_k) \nonumber \\
&=&\sum_{i_1,\dots,i_\mu}\epsilon_{i_1\dots i_\mu}^2
\prod_{k=1}^\mu\Gamma(i_k)\Gamma(\alpha+i_k) \nonumber \\
&=&\mu!\prod_{k=1}^\mu\Gamma(k)\Gamma(\alpha+k)\;.
\end{eqnarray}
Elaborations of this calculation lead to the following formulae:
\begin{eqnarray}
\sum_{i=1}^\mu
\int q_i^4 Q(\mathbf{q})\,d\mathbf{q} &=&
\overline{Q} \sum_{k=0}^{\mu-1}\frac{I_{kk}^4(\alpha)}
{\Gamma(k+1)\Gamma(\alpha+k+1)}\;,
\label{S1}\\
\sum_{i\neq j=1}^\mu
\int q_i^2q_j^2 Q(\mathbf{q})\,d\mathbf{q} &=&
\overline{Q} \sum_{k,l=0}^{\mu-1}
\frac{I_{kk}^2(\alpha)I_{ll}^2(\alpha)-\left[I_{kl}^2(\alpha)\right]^2}
{\Gamma(k+1)\Gamma(\alpha+k+1)\Gamma(l+1)\Gamma(\alpha+l+1)}\;.
\label{S2}
\end{eqnarray}
Here
\begin{eqnarray}
I_{kl}^j(\alpha) &\equiv&
\int_0^{\infty}dq\,e^{-q}q^{\alpha+j}r^\alpha_k(q)r^\alpha_l(q)\nonumber\\
&=& \Gamma(k+1)\sum_{i,r=0}^j(-1)^{i+r}{\,j\,\choose i}
{\,j\,\choose r}\Gamma(\alpha+j+l-r+1)l(l-1)\dots(l-r+1)\delta_{l-r,k-i}\;,
\end{eqnarray}
where the final form follows from Eqs.~(\ref{L3}) and (\ref{L1}).
Evaluating the sums in Eqs.~(\ref{S1}) and (\ref{S2}) leads to the
simplification
\begin{eqnarray}
\sum_{i,j=1}^\mu\int q_i^2q_j^2 Q(\mathbf{q})\,d\mathbf{q} &=&
\overline{Q}\,\mu\nu\big[(\mu+\nu)(\mu^2+\nu^2+5\mu\nu+5)+
(\mu-1)(\nu-1)(\mu+\nu-1)(\mu+\nu-2)\big]\;.
\end{eqnarray}

Using Eqs.~(\ref{R2}) and (\ref{QtoP}), we now obtain
\begin{equation}
\langle R^2\rangle=
\frac{(\mu+\nu)(\mu^2+\nu^2+5\mu\nu+5)+(\mu-1)(\nu-1)(\mu+\nu-1)(\mu+\nu-2)}
{(\mu\nu+3)(\mu\nu+2)(\mu\nu+1)}\;.
\end{equation}
The variance is then given by
\begin{equation}
\langle\!\langle R^2\rangle\!\rangle\equiv
\langle R^2\rangle-\langle R\rangle^2=
\frac{2(\mu^2-1)(\nu^2-1)}{(\mu\nu+3)(\mu\nu+2)(\mu\nu+1)^2}\;.
\end{equation}
Using the same methods, one can also derive an expression for the
third cumulant.  Due to the complexity of these calculations, we only
state the final result:
\begin{equation}
\langle\!\langle R^3\rangle\!\rangle \equiv
\langle R^3\rangle-3\langle R\rangle\langle R^2\rangle+2\langle R\rangle^3
=\frac{8(\mu^2-1)(\nu^2-1)(\mu+\nu)(\mu\nu-5)}
{(\mu\nu+5)(\mu\nu+4)(\mu\nu+3)(\mu\nu+2)(\mu\nu+1)^3}\;.
\end{equation}

\begin{figure}[t]
\includegraphics[scale=0.9]{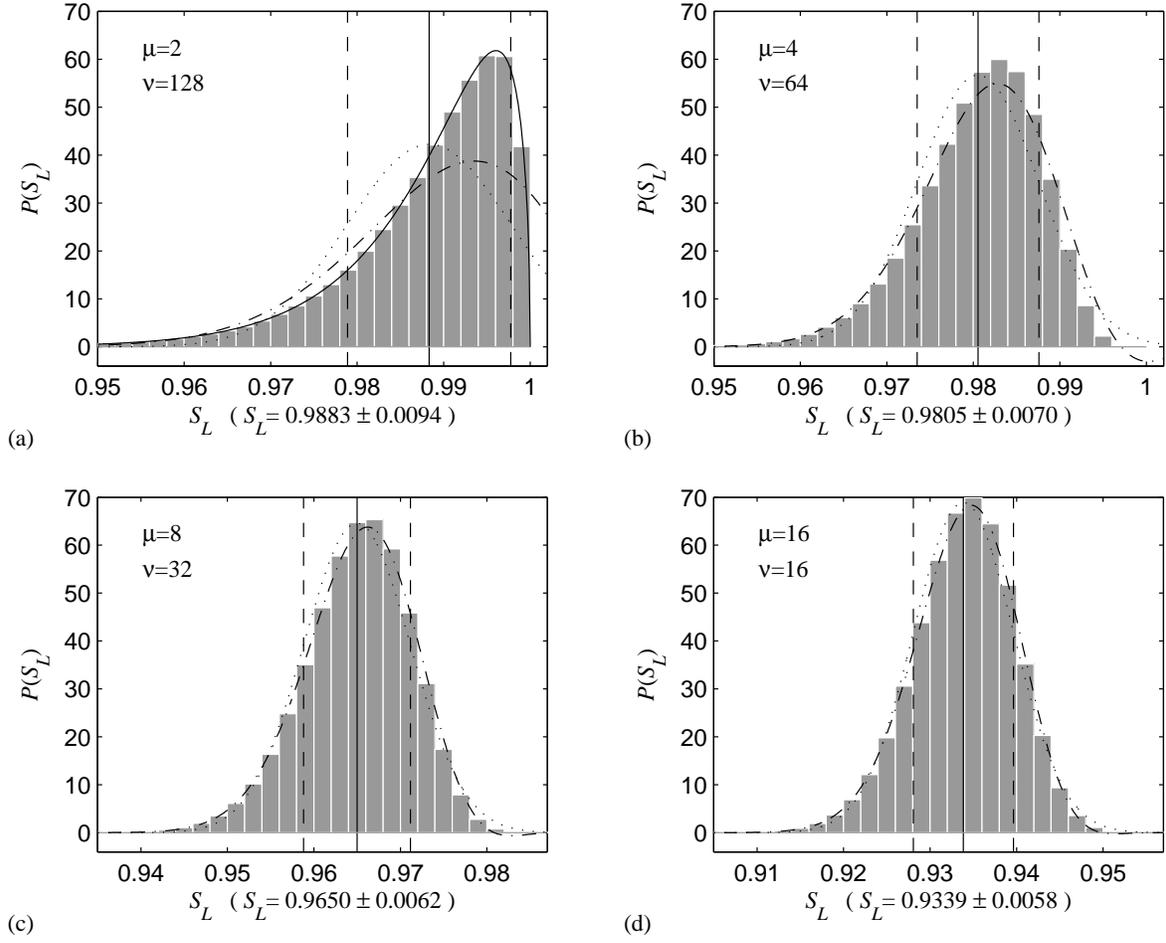}
\caption{Histograms for the probability distribution $P(S_L)$,
numerically calculated from 1 million random states for the several
subsystem dimensions $\mu$ and $\nu$ possible in a 256-dimensional
Hilbert space.  The exact means (vertical solid line) and standard
deviations (vertical dashed line), along with Gaussian (dotted curve)
and Airy function (dash-dotted curve) approximations to the
distribution are also plotted.  For $\mu=2$, the exact
distribution~(\ref{exacttwo}) is plotted as the solid curve.  Notice
that for $\mu$ not too close to 2, the Gaussian approximation is
sufficient and, as noted in the text, the linear entropy of a typical
state, though close to maximal, is localized away from maximal.}
\label{fig2}
\end{figure}

Translating our results to the linear entropy, we have
\begin{eqnarray}
\langle\!\langle S_L\rangle\!\rangle &\equiv&
\langle S_L\rangle=\beta\frac{(\mu-1)(\nu-1)}{\mu\nu+1} \quad(\equiv a)\;,\\
\langle\!\langle S_L^2\rangle\!\rangle &=&
\beta^2\langle\!\langle R^2\rangle\!\rangle \qquad\qquad\quad(\equiv b)\;,\\
\langle\!\langle S_L^3\rangle\!\rangle &=&
-\beta^3\langle\!\langle R^3\rangle\!\rangle \;\quad\qquad\quad(\equiv c)\;.
\end{eqnarray}
These can be used in an approximation to the cumulant generating
function and, hence, to the probability distribution itself:
\begin{eqnarray}
P(S_L)dS_L&\approx&
\frac{1}{2\pi}\int_{-\infty}^{\infty}d\omega\,
\exp\left[-iS_L\omega +ai\omega+b(i\omega)^2/2!+c(i\omega)^3/3!\right]dS_L
\nonumber \\
&=&|2/c|^{1/3}
\exp\left[\frac{b^3}{3c^2}+\frac{b(S_L-a)}{c}\right]\mathrm{Ai}
\left[(2/c)^{1/3}\left(S_L-a+\frac{b^2}{2c}\right)\right]dS_L\;.
\label{airy}
\end{eqnarray}
When the overall dimension $\mu\nu$ is large, the standard deviation
of the linear entropy is given approximately by
\begin{equation}
\langle\!\langle S_L^2\rangle\!\rangle^{1/2}\approx
{\mu+1\over\mu\nu+1}\sqrt{{2\over\mu^2-1}}
\left(1-{5\over2\mu\nu}-{1\over2\nu^2}\right)\;.
\end{equation}
Comparing this with the average linear entropy shows that the
bipartite entanglement of a typical pure state, though close to
maximal, is nonetheless localized away from maximal as long as
$\langle\!\langle S_L^2\rangle\!\rangle^{1/2}$ is somewhat smaller
than $(\mu+1)/(\mu\nu+1)$, i.e., as long as $\mu$ is somewhat larger
than 2.

In Fig. \ref{fig2} we display numerical calculations of $P(S_L)$ for
the several ways of dividing a 256-dimensional Hilbert space into two
subsystems.  These numerical calculations used 1 million random
states.  We also plot the means (vertical solid line), standard
deviations (vertical dashed line), the Airy function
approximation~(\ref{airy}) (dash-dotted curve), and the Gaussian
approximation (dotted curve).  For the special case $\mu=2$, the
exact probability distribution is drawn (solid curve):
\begin{equation}
P(S_L)\,dS_L=
\frac{2\Gamma(\nu+1/2)}{\sqrt{\pi}\,\Gamma(\nu-1)}
\sqrt{1-S_L}{S_L}^{\!\nu-2}dS_L\qquad(\mu=2)\;.
\label{exacttwo}
\end{equation}
Note that the distributions are highly localized and that for $\mu$
somewhat larger than 2 in a high-dimensional overall space, the
Gaussian approximation is sufficient.

\begin{figure}[t]
\includegraphics[scale=1]{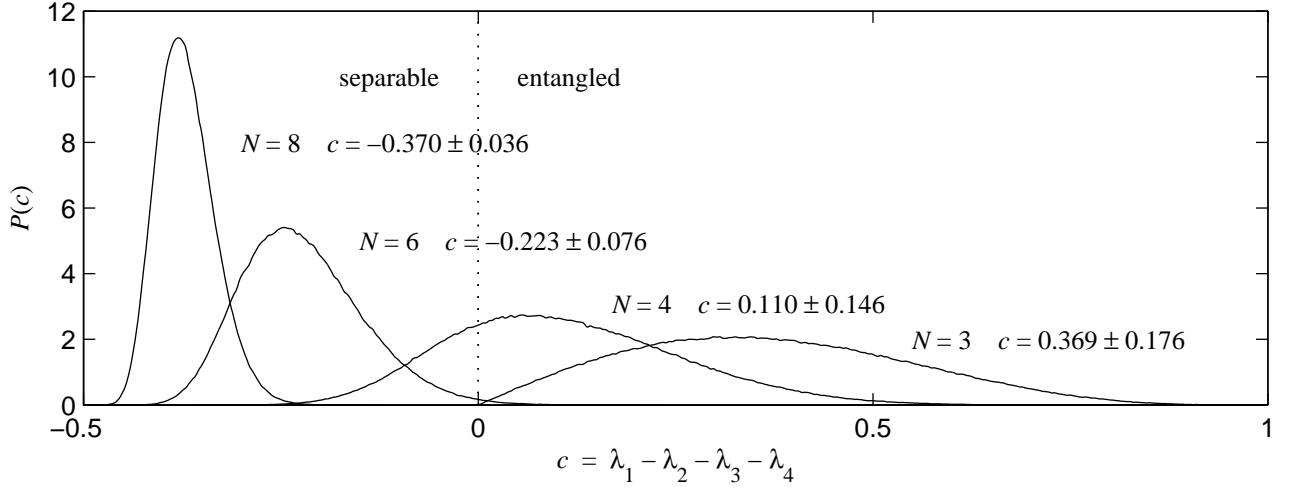}
\caption{Pairwise mixed-state entanglement in random multi-qubit pure
states with a total of 3, 4, 6, and 8 qubits.  The plots show the
probability distribution of the quantity
$c(\hat\rho)=\lambda_1-\lambda_2-\lambda_3-\lambda_4$.  The
concurrence is $C(\hat\rho)=\max\{0,c(\hat\rho)\}$.  The means and
standard deviations of the distributions are also given.  The
approximate probability to encounter pairwise entanglement between a
particular pair of qubits in states with 3, 4, 6, and 8 total qubits
is 1, 0.76, 0.006 and 0, respectively.  The distributions were
numerically calculated using 1 million random states.} \label{fig3}
\end{figure}

\subsection{Pairwise mixed-state entanglement}
\label{sec3b}

A numerical study of pairwise (mixed-state) entanglement in
multi-qubit systems has already been published \cite{kendon}, making
the results in this section somewhat redundant.  Our choice for
an entanglement measure is the {\it concurrence\/} \cite{wootters} of
a two-qubit density operator $\hat\rho$, given by
\begin{equation}
C(\hat\rho)\equiv\max\{0,\lambda_1-\lambda_2-\lambda_3-\lambda_4\}
\end{equation}
where $\lambda_1\geq\lambda_2\geq\lambda_3\geq\lambda_4$ are the
square roots of the eigenvalues of
$\hat\rho(\hat\sigma_y\otimes\hat\sigma_y)
\hat\rho^*(\hat\sigma_y\otimes\hat\sigma_y)$.  The complex
conjugation is taken in the standard qubit basis. The concurrence
takes values in the range $[0,1]$, with a pair of qubits being
entangled if and only if $C(\hat\rho)>0$.

The concurrence provides an explicit formula for the entanglement of
formation
\begin{equation}
E_f(\hat\rho)\equiv \inf \sum_j p_j S(\psi_j)\;,
\end{equation}
where the infimum is taken over all pure-state decompositions
$\hat\rho=\sum_j p_j \ket{\psi_j}\bra{\psi_j}$, and $S(\psi_j)$
is the subsystem von Neumann entropy of the bipartite pure state
$\psi_j$ .  In the case of two qubits,
\begin{equation}
E_f(\hat\rho)={\cal E}\big(C(\hat\rho)\big)\;,
\end{equation}
where ${\cal E}$ is defined in terms of the binary entropy function
$h(x)=-x\log x-(1-x)\log(1-x)$ by
\begin{equation}
{\cal E}(C)\equiv h\!\left(\frac{1+\sqrt{1-C^2}}{2}\right)\;.
\end{equation}
The function ${\cal E}(C)$ is monotonically increasing for $0\leq
C\leq 1$, and hence the concurrence is a good measure of entanglement
in its own right.

To apply the concurrence as a measure of pairwise entanglement in
$N$-qubit pure states, we first trace out $N-2$ of the qubits and
then use the above formulae on the resulting two-qubit density
operator. In Fig. \ref{fig3} we have plotted the probability
distribution for the quantity
\begin{equation}
c(\hat\rho)\equiv\lambda_1-\lambda_2-\lambda_3-\lambda_4
\quad(-1/2\leq c\leq 1)\;,
\label{crho}
\end{equation}
when the $N$-qubit pure states are sampled from the Haar distribution
for $N=3$, 4, 6, and 8.  Notice that the probability of finding
pairwise entanglement between any pair of qubits decreases rapidly as
$N$ increases.  In contrast, the preceding subsection shows that as
$N$ increases, the bipartite entanglement of a typical state is close
to maximal no matter how the overall system is sliced into two parts.
Taken together, these results mean that the entanglement in a typical
state of many qubits is mainly multipartite entanglement shared among
many of the qubits, rather than pairwise entanglement.

\begin{figure}[t]
\includegraphics[scale=1]{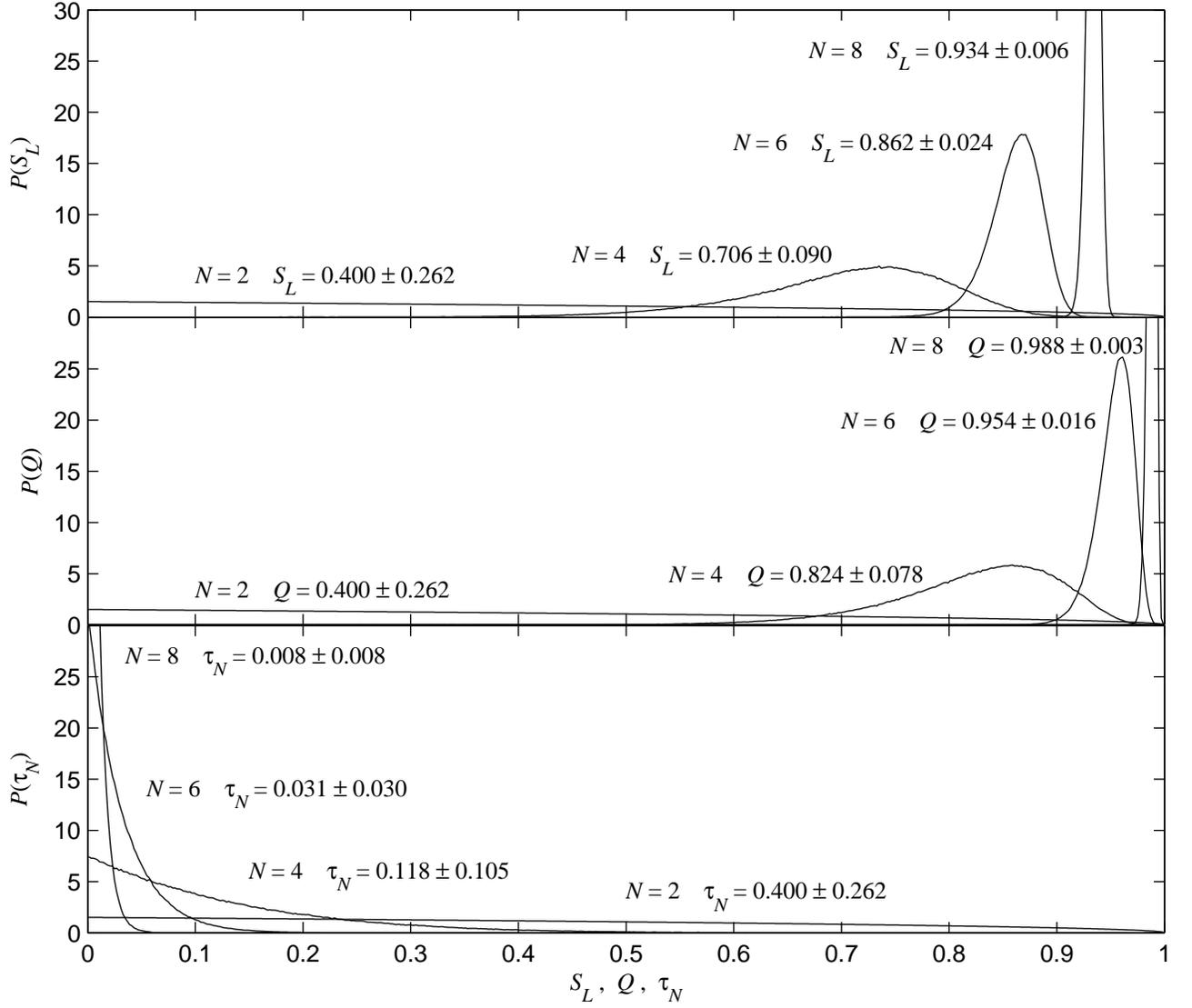}
\caption{Multipartite entanglement in random multi-qubit pure states
with a total of 2, 4, 6, and 8 qubits.  The probability distributions
are for the multipartite entanglement measures $Q$ and $\tau_{N}$.
For comparison, the distributions for the subsystem linear entropy
$S_L$ are also plotted in the case of equal subsystem dimensions
($\mu=\nu=\sqrt{D}$).  The analytic values of the means and standard
deviations are also shown.  The distributions are based on numerical
calculations using 1 million random states.} \label{fig4}
\end{figure}

\subsection{Multipartite entanglement}
\label{sec3c}

We now investigate two proposals for a measure of multipartite
entanglement, the measure $Q$ of Meyer and Wallach \cite{meyer} and
the $n$-tangle of Wong and Christensen \cite{wong}.  In general, as
the number of subsystems increases, an exponential number of
independent measures is needed to quantify fully the amount
entanglement in the multipartite system.  Consequently, neither of
the following entanglement measures can be thought of as unique.
Different measures capture different aspects of multipartite
entanglement.

The Meyer-Wallach measure, $Q(\psi)$, which can only be applied to
multi-qubit pure states, is defined as follows.  For each
$\alpha=1,\dots,N$ and $b\in\{0,1\}$, we define the linear map
$\imath_\alpha(b):(\mathbb{C}^2)^{\otimes
N}\rightarrow(\mathbb{C}^2)^{\otimes N-1}$ through its action on the
product basis,
\begin{equation}
\imath_\alpha(b)\ket{x_1}\otimes\dots\otimes\ket{x_N}=
\delta_{bx_\alpha}\ket{x_1}\otimes\dots\otimes\ket{x_{\alpha-1}}\otimes
\ket{x_{\alpha+1}}\otimes\dots\otimes\ket{x_N}\;.
\end{equation}
Next let
\begin{equation}
D(\psi,\phi)=\sum_{i<j}|\psi_i\phi_j-\psi_j\phi_i|^2\;,
\end{equation}
be the square of the wedge product of two vectors $\ket{\psi}$ and
$\ket{\phi}$, where the $\psi_j$ are the coefficients of the state
$\ket{\psi}$ in the product basis,
\begin{equation}
\ket{\psi}=\sum_{j=1}^{2^m} \psi_j \ket{x_1}\otimes\dots\otimes\ket{x_m}\;,
\qquad
j=x_1\ldots x_m\!\cdot\!0=\sum_{l=1}^m x_l 2^{m-l}\;.
\end{equation}
The Meyer-Wallach entanglement measure is then
\begin{equation}
Q(\psi)\equiv
\frac{4}{N}\sum_{\alpha=1}^N
D\big(\imath_\alpha(0)\psi,\imath_\alpha(1)\psi\big)\;.
\end{equation}
Meyer and Wallach have shown that $Q$ is invariant under local
unitary transformations and that $0\leq Q\leq 1$, with $Q(\psi)=0$ if
and only if $\ket\psi$ is a product state. It was recently shown by
Brennen \cite{brennen} that $Q$ is simply the average subsystem linear
entropy of the constituent qubits:
\begin{equation}
Q(\psi)=2\left(1-\frac{1}{N}\sum_{k=1}^N\tr\,\hat{\rho}_k^2\right). \label{brennenformula}
\end{equation}
Here $\hat{\rho}_k$ is the density operator for the $k$-th qubit
after tracing out the rest. Hence we should expect this measure to
agree qualitatively with the subsystem entropies.

\begin{figure}[p]
\includegraphics[scale=0.88]{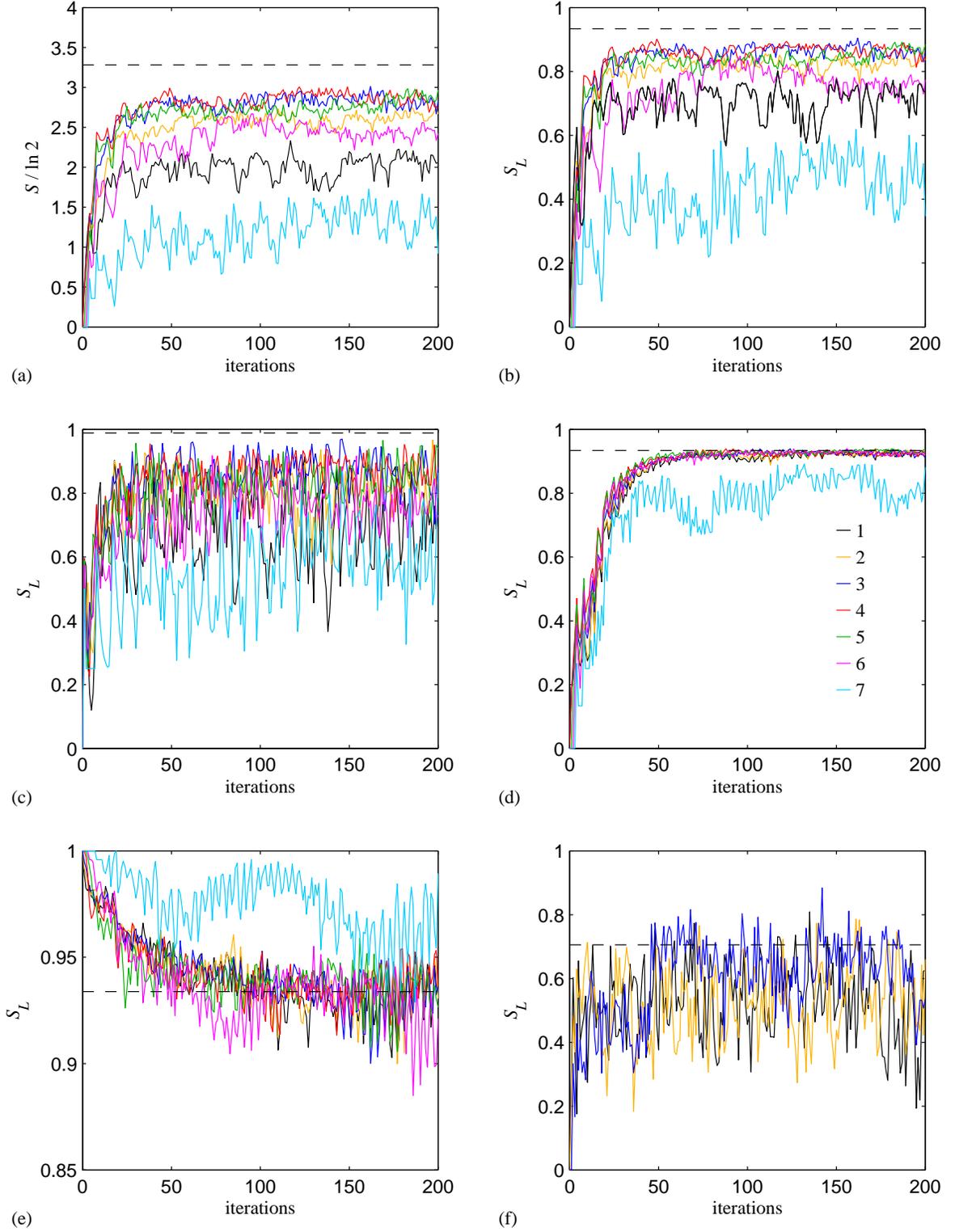}
\caption{Dynamical behavior of the subsystem entropy.  (a) and (b)
plot the von Neumann entropy and the linear entropy, respectively,
when $N=8$, the partition divides the four rightmost qubits from the
four leftmost, and the initial state is $\ket{00000000}$.  (c) is the
same as (b), except that the partition divides the single rightmost
qubit from the others.  (d) and (e) are also the same as (b), but
with initial state $\ket{00100111}$ in (d) and maximally entangled
initial state~(\ref{4state}) in (e).  (f) shows the case of $N=4$
qubits, a partition that separates the two leftmost qubits from the
two rightmost, and an initial state $\ket{0000}$.  The different
quantum baker's maps are colored according to the key in~(d).  The
dashed lines show the average entanglement predicted for random
states.} \label{fig5}
\end{figure}

Using, for example, the Hurwitz parametrization
\cite{hurwitz,pozniak} of a random unit vector in $\mathbb{C}^D$,
where the dimension $D=2^N$, one can calculate the mean and variance
of $Q$ for random states:
\begin{eqnarray}
&&\langle Q \rangle=\frac{D-2}{D+1}\;, \\
&&\langle\!\langle Q^2\rangle\!\rangle=
\langle Q^2 \rangle - \langle Q \rangle^2=
\frac{6(D-4)}{(D+3)(D+2)(D+1)N}+\frac{18D}{(D+3)(D+2)(D+1)^2}\;.
\end{eqnarray}
The mean was also calculated independently in \cite{weinstein}, and
given the relationship (\ref{brennenformula}), it can be easily
checked using Lubkin's formula for the average purity (\ref{lubkinformula}).
For large $N$, the mean and standard deviation are given
approximately by $\langle Q\rangle\approx1-3/D$ and $\langle\!\langle
Q^2\rangle\!\rangle^{1/2} \approx(1/D)\sqrt{6/N}$, indicating that
the Meyer-Wallach entanglement of a typical state is very nearly
maximal, but suggesting that this entanglement is weakly localized
just below maximal.

In the case of two qubits, the square of the concurrence is often
referred to as the {\it tangle}.  A generalization of the tangle to
three qubits was defined by Coffman {\it et al.} \cite{coffman}. Wong
and Christensen proposed another generalization valid for an
arbitrary even number of qubits.  For pure states of $N$ (even)
qubits, it is defined as
\begin{equation}
\tau_{N}^{\phantom{2}}(\psi) \equiv
|\langle\psi|\sigma_y^{\otimes N}|\psi^*\rangle|^2
\qquad
(N \textrm{ even})\;.
\end{equation}
Wong and Christensen were able to show that $\tau_{N}$ is an
entanglement monotone and also to generalize its definition to mixed
states. From the above definition, it is easy to see that
$0\leq\tau_{N}^{\phantom{2}}\leq 1$.  A cat state,
$(|0\rangle^{\otimes N}+|1\rangle^{\otimes N})/\sqrt2$, has maximal
entanglement by this measure.

One can calculate the mean and variance of $\tau_{N}$ for random
states:
\begin{eqnarray}
&&\langle \tau_{N}^{\phantom{2}} \rangle = \frac{2}{D+1}\;, \\
&&\langle\!\langle \tau_{N}^2\rangle\!\rangle=
\langle \tau_{N}^2 \rangle - \langle \tau_{N}^{\phantom{2}} \rangle^2
=\frac{4(D-1)}{(D+3)(D+1)^2}\;.
\end{eqnarray}
For large $N$, both the mean and the standard deviation of $\tau_N$
are given approximately by $2/D$, indicating that a typical state
does not have whatever sort of multipartite entanglement is
characterized by $\tau_N$.

In Fig.~\ref{fig4} we plot the probability distributions of $Q$ and
$\tau_{N}$ after sampling 1 million random multi-qubit pure states
with a total of 2, 4, 6, and 8 qubits.  Also plotted, for comparison,
are the corresponding distributions for the subsystem linear entropy
$S_L$, with the two subsystem dimensions chosen to be equal.  When
$N=2$, the three different measures are equivalent, each having the
distribution $P(E)\,dE=(3/2)\sqrt{1-E}\,dE$ ($E=S_L,Q,\tau_N$). Notice,
however, that the behaviors of the two multipartite measures diverge
dramatically as we increase the total number of qubits.  According to
the Meyer-Wallach measure, multipartite entanglement increases as we
increase the number of qubits.  This agrees with the bipartite
measure $S_L$, the only difference being that the Meyer-Wallach
entanglement of a typical state is closer to maximal than the linear
entropy as $N$ gets large.  In contrast, the $n$-tangle of a typical
state decreases as $N$ increases.  As noted above, $\tau_{N}$ seems
to describe some sort of multipartite entanglement that becomes rare
as the number of qubits increases.

\section{Entanglement production in the quantum baker's maps}
\label{sec4}

\begin{figure}[p]
\includegraphics[scale=0.9]{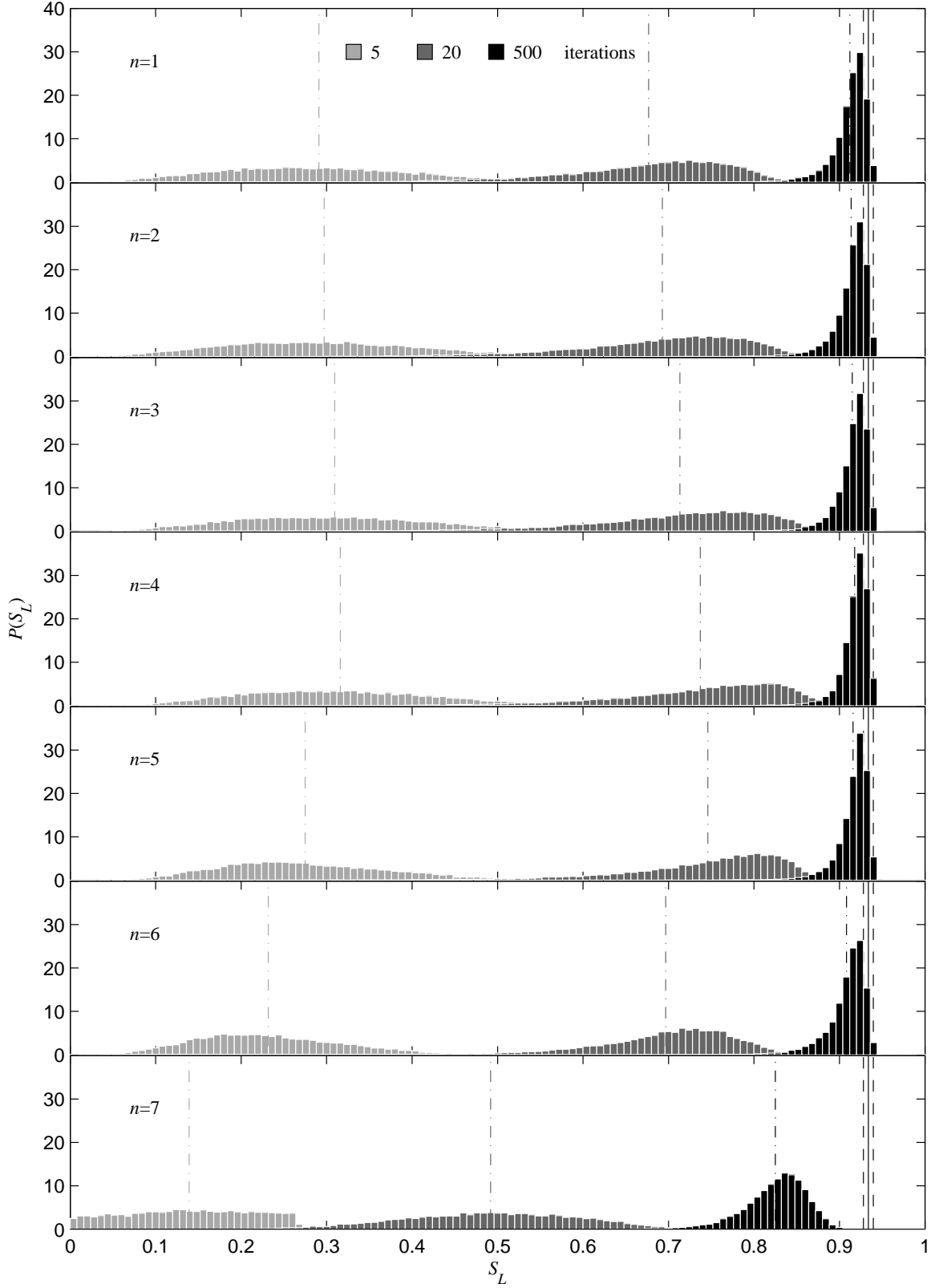}
\caption{Distributions for the subsystem linear entropy, relative to
a partition that divides the four least significant qubits from the
four most significant, after 5, 20, and 500 iterations of the quantum
baker's maps starting with a uniform distribution of $\states$
product states (delta function centered at zero).  A total of $N=8$
qubits is used.  The dashed-dotted lines are the means of the
distributions, and the mean and standard deviation for random states
are the solid and dashed lines, respectively.} \label{fig6}
\end{figure}

\begin{figure}[t]
\includegraphics[scale=0.95]{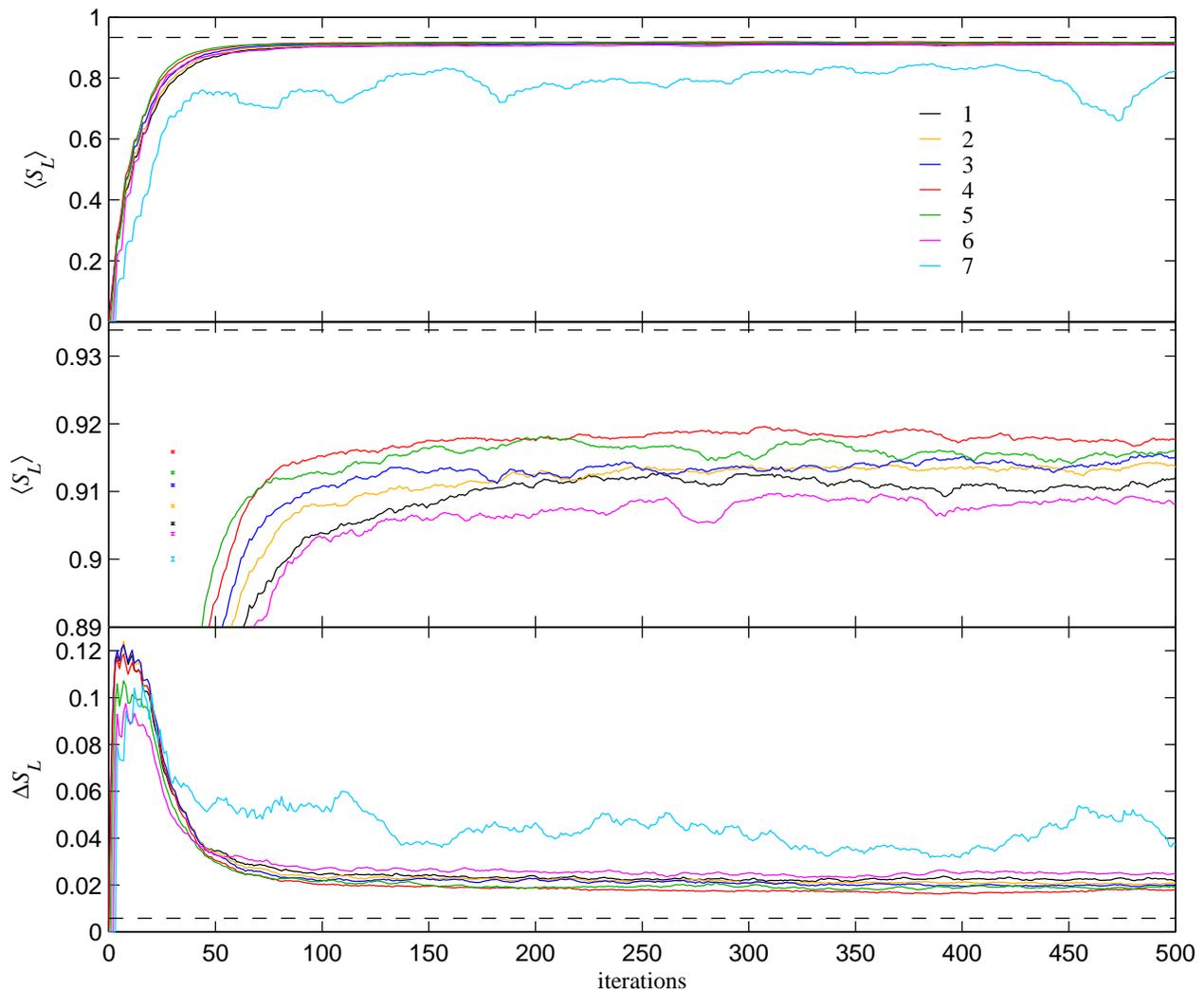}
\caption{Evolution of the means and standard deviations for the
subsystem linear entropy under the same conditions as in
Fig.~\ref{fig6}.  The dashed lines are the mean and standard
deviation for random states.   The middle, magnified plot displays
the ranking of the baker's maps in terms of entangling power, i.e.,
4,5,3,2,1,6,7.  Error bars ($2\Delta S_L/\sqrt{\states}$) are
included in the middle plot to indicate the expected sampling error
for the means at saturation; these error bars show that the rankings
in terms of entangling power are real, not statistical artifacts.}
\label{fig7}
\end{figure}

\begin{figure}[t]
\includegraphics[scale=0.95]{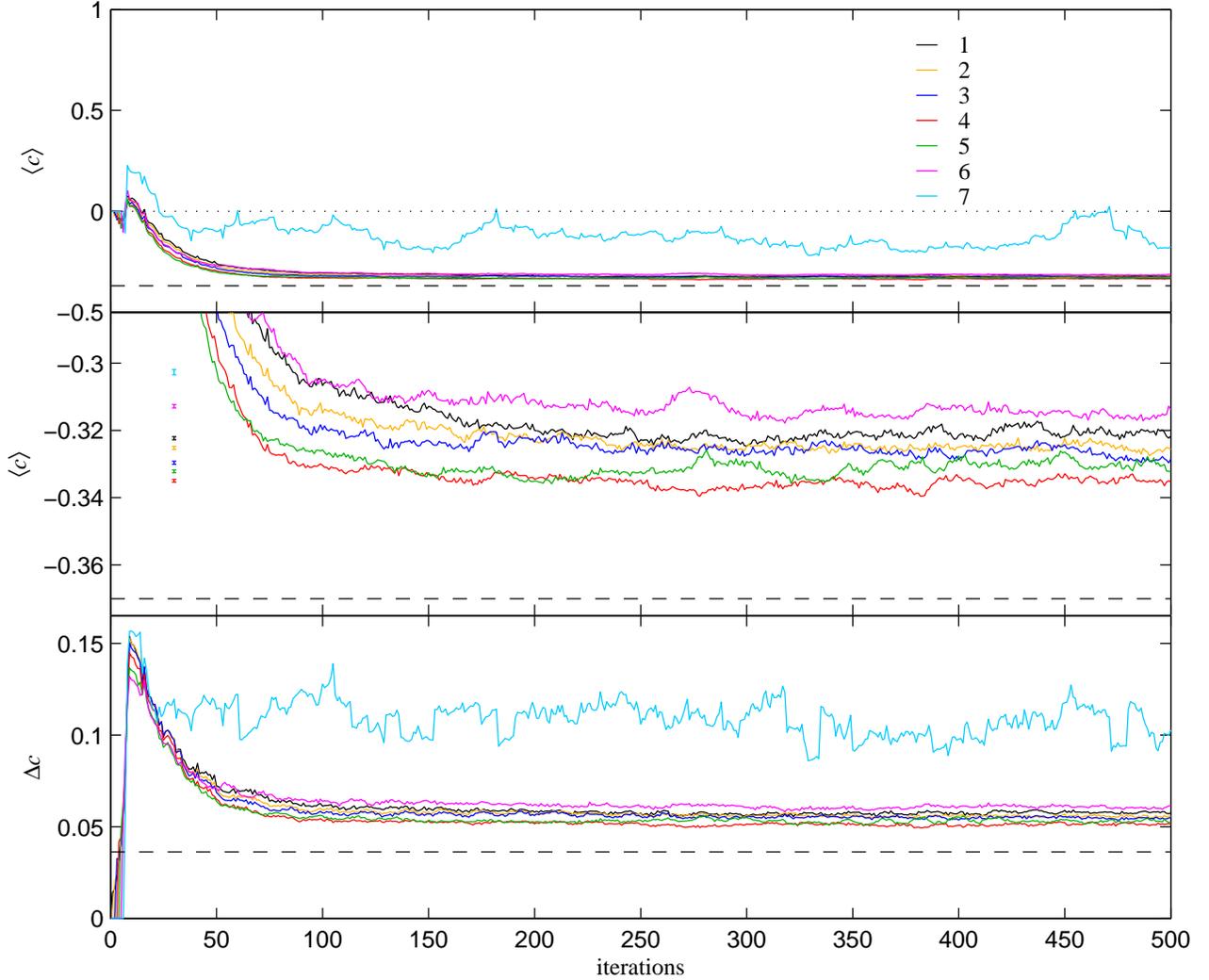}
\caption{Evolution of the means and standard deviations for the
quantity $c$ of Eq.~(\ref{crho}), where all but the single leftmost
and rightmost qubits are traced out.  The concurrence for the
remaining two qubits is $C=\max\{0,c\}$.  The averages are taken with
respect to $\states$ uniformly distributed initial product states for
$N=8$ qubits.  The dashed lines are the numerically calculated mean
and standard deviation for random states.  Error bars ($2\Delta
c/\sqrt{\states}$) are included in the middle plot to indicate the
expected sampling error for the means at saturation.} \label{fig8}
\end{figure}

\begin{figure}[t]
\includegraphics[scale=0.95]{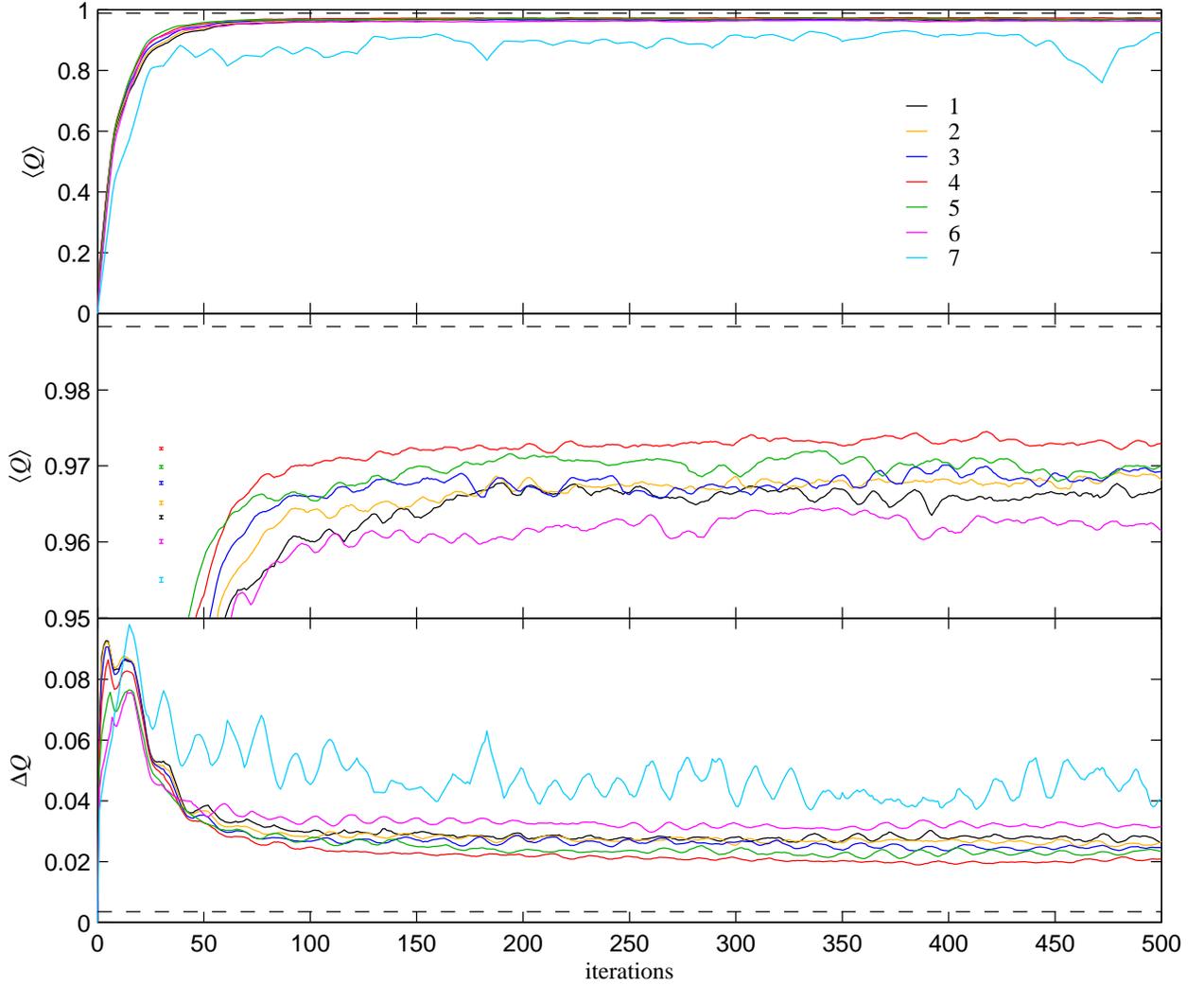}
\caption{Evolution of the means and standard deviations for the
Meyer-Wallach multipartite entanglement measure $Q$.  The averages
are taken with respect to $\states$ uniformly distributed initial
product states for $N=8$ qubits.  The dashed lines are the
numerically calculated mean and standard deviation for random states.
Error bars ($2\Delta Q/\sqrt{\states}$) are included in the middle
plot to indicate the expected sampling error for the means at
saturation.} \label{fig9}
\end{figure}

\begin{figure}[t]
\includegraphics[scale=0.95]{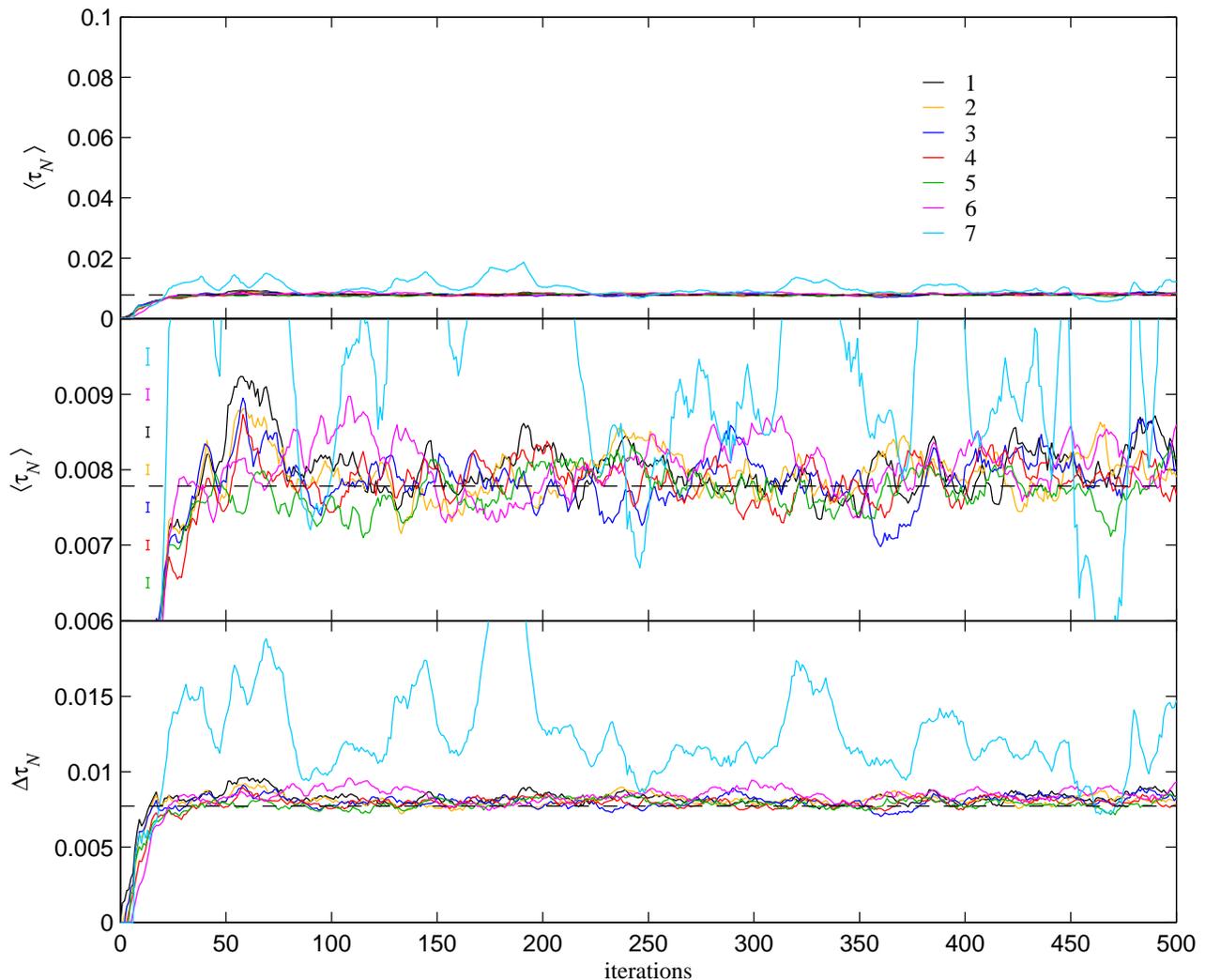}
\caption{Evolution of the means and standard deviations for the
Wong-Christensen tangle $\tau_{N}$.  The averages are taken with
respect to $\states$ uniformly distributed initial product states for
$N=8$ qubits.  The dashed lines are the numerically calculated mean
and standard deviation for random states.  Error bars
($2\Delta\tau_N/\sqrt{\states}$) are included in the middle plot to
indicate the expected sampling error for the means at saturation.}
\label{fig10}
\end{figure}

\begin{figure}[t]
\includegraphics[scale=0.95]{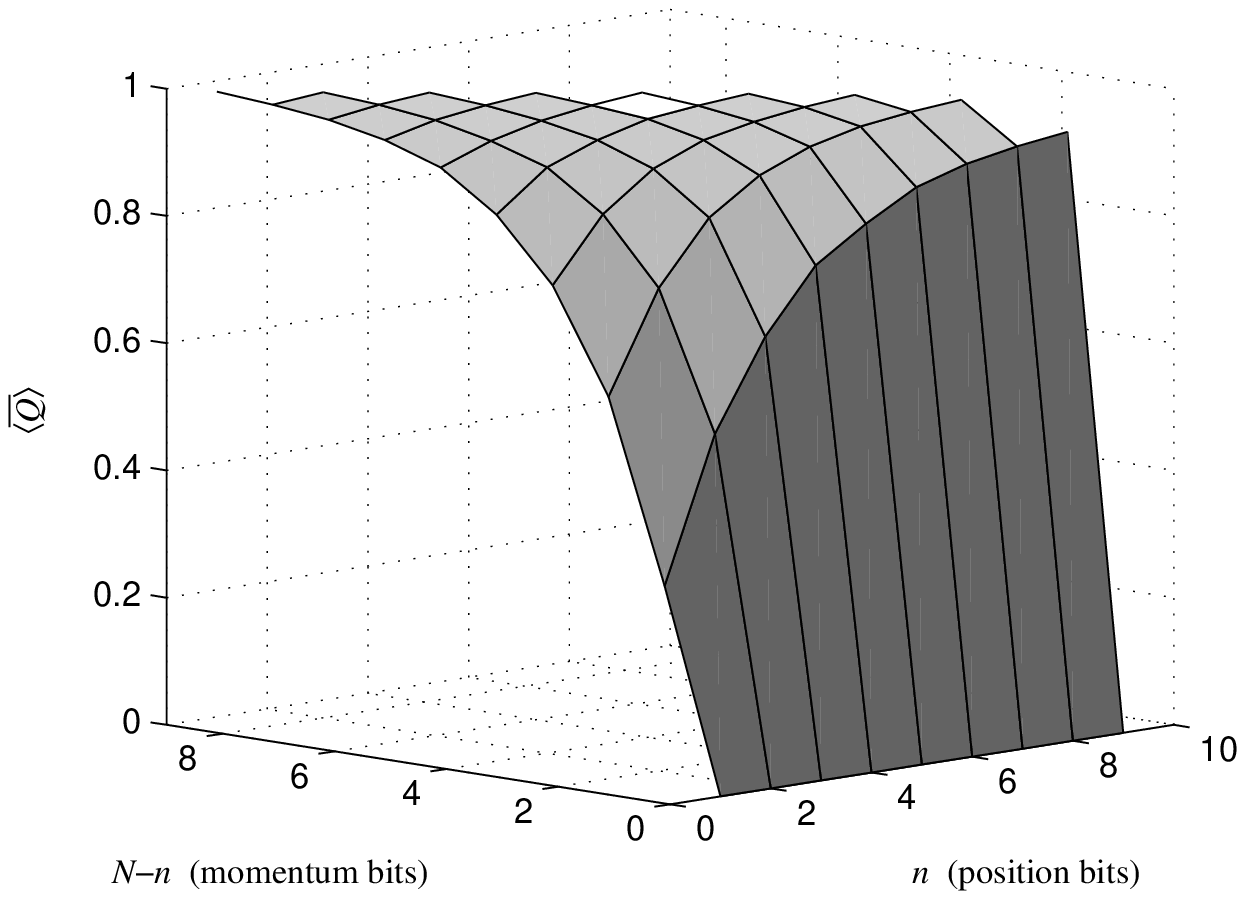}
\caption{Saturation value $\overline{\langle Q\rangle}$ of the
Meyer-Wallach entanglement measure for all quantum baker's maps up to
a total of $N=10$ qubits. This quantity is approximated by taking a
time average over the $(512k)$-th iterates ($k=1,\dots,100$) and
using $4\,000$ uniformly distributed initial product states.}
\label{fig11}
\end{figure}

We now report the results of a detailed numerical study of
entanglement production for the quantum baker's maps.  All of the
results reported in this section are for the 8-qubit baker's maps
(except for Fig.~\ref{fig5}(f), which displays results for $N=4$
qubits).  In the figures, the maps for different values of $n$ are
distinguished by a color coding given in Fig.~\ref{fig5}(d).  The
extreme map $\hat B_{N,N}$ is not included in the plots because it
does not produce any entanglement.

First consider the bipartite measures of entanglement.  In
Fig.~\ref{fig5} we plot the dynamical behavior of these measures for
several interesting initial states.  Figure~\ref{fig5}(a) shows the
behavior of the subsystem von Neumann entropy $S$ when $N=8$ and the
bipartite division is between the four least significant qubits
(rightmost) and the four most significant (leftmost); in terms of the
phase-space description of the baker's maps, we are considering
entanglement between the fine and coarse position scales on phase
space.  The initial state in Fig.~\ref{fig5}(a) is the product state
$\ket{00000000}$.  Note the ranking among the different quantum
baker's maps.  The maps $\hat B_{8,4}$ (red) and $\hat B_{8,3}$
(dark blue) achieve
saturation values closest to that for typical states (dashed line).
The original Balazs-Voros-Saraceno quantization has $n=1$ (black).
Figure~\ref{fig5}(b) is the same as (a), except now the linear
entropy $S_L$ is our entanglement measure (as it remains in the
remaining parts of Fig.~\ref{fig5}).  The two different definitions
of entropy show the same qualitative behavior.  In
Fig.~\ref{fig5}(c), we again use linear entropy, but switch to a
bipartite division that separates the single rightmost qubit from the
rest.  The different saturation levels attained by the different
quantum baker's maps, though less intelligible, are still
discernible.  If we return to our original partition and change the
initial state to $\ket{00100111}$, however, the differences in the
saturation levels disappear almost altogether, as is shown in
Fig.~\ref{fig5}(d).  The quantum baker's map with $n=7$ stands out in
all cases, saturating at a value well below the other maps; we should
recall that the map $\hat B_{8,7}$ is closest to the map $\hat
B_{8,8}$, which does not entangle at all.  In Fig.~\ref{fig5}(e),
again using the original partition, we start in the entangled state
\begin{equation}
{1\over4}\sum_{x_1,x_2,x_3,x_4}\ket{x_1x_2x_3x_4x_1x_2x_3x_4}\;,
\label{4state}
\end{equation}
which is maximally entangled with respect to the original partition.
In this case the maximal initial level of entanglement is destroyed
by the quantum baker's maps.  Figure~\ref{fig5}(f) shows the case
with $N=4$ qubits, a partition separating the two leftmost qubits
from the two rightmost, and an initial state of $\ket{0000}$.

In view of the variety of behaviors exhibited by subtle differences
in the initial states [compare Fig.~\ref{fig5}(b) to \ref{fig5}(d)],
if we are to study the intrinsic entangling properties of the quantum
baker's maps and not properties conditioned on a particular initial
state, we need an approach that treats all initial states of a
certain type on the same footing. Such neutrality can be achieved by
taking an average over, for example, the set of all product states.
This approach was used to define the {\it entangling power\/}
\cite{zanardi} of a unitary operator, and we adhere to it for the
remainder of this section.

Consider starting with a uniform distribution of product pure states,
each member being a tensor product of $N=8$ random single-qubit
states.  We now ``bake'' entanglement with the seven quantum baker's
maps.  The remaining figures plot the amount of baked entanglement as
quantified by the several entanglement measures discussed in
Sec.~\ref{sec3}.

Figure~\ref{fig6} plots the distributions of the subsystem linear
entropy, relative to a partition that divides the four least
significant qubits from the four most significant, after 5, 20, and
500 iterations of the quantum baker's maps.  All the distributions
start in a delta function centered at zero entanglement, but quickly
spread while advancing toward the value predicted for typical states.
The mean and standard deviation for random states are the solid and
dashed lines, respectively.  At 500 iterations, well after saturation
($\sim 100$), the mean entanglement for our quantum-baked states
(dash-dotted lines) fall short of the value predicted for random
states.  Notice that the variances are also still quite large.  For
clarity, the means and standard deviations are plotted separately in
Fig.~\ref{fig7}.  The ranking of the different quantum baker's maps
in terms of entangling power---4,5,3,2,1,6,7---is now only apparent
after magnification at the saturation level (middle plot).  It was
found that this ranking is preserved when the partition is changed.
If the total number of qubits is varied, similar behavior also
results, and we conclude that the quantum baker's maps are, in
general, good generators of bipartite entanglement, with the highest
levels produced when $n$ is roughly midway between $N$ and $1$.  A
total of $\states$ initial states were used in this simulation.
The same number is used for those that follow.

We next consider the pairwise entanglement as given by the
concurrence, or more specifically, by the quantity $c(\hat\rho)$ of
Eq.~(\ref{crho}).  We only display results for the case where all but
the single leftmost and rightmost qubits are traced out.  The other
cases are similar.  Again, starting in a uniform distribution of
$\states$ product states (delta function centered at $c=0$), we bake
entanglement into our states.  In this case, however, the pairwise
entanglement is not maintained.  After a short rise, $c(\hat\rho)$
quickly falls to negative values, retreating toward the mean
numerically calculated for random states, as shown in
Fig.~\ref{fig8}.  Not surprisingly, the quantum baker's maps respect
the rarity of pairwise entanglement in multi-qubit states.  The
ranking among the different quantum baker's maps is also preserved.

In Figs.~\ref{fig9} and \ref{fig10}, we plot the corresponding
evolutions for multipartite entanglement measures, the Meyer-Wallach
measure $Q$ and Wong-Christensen $\tau_{N}$.  The means and standard
deviations of $Q$ behave similarly to the bipartite measures of
entanglement.  The ranking among the different quantum baker's maps
is again preserved.  This is not surprising given the relationship
(\ref{brennenformula}).  In the case of $\tau_{N}$, however, it is
difficult to discern any useful information, presumably because this
measure only describes entanglement of a very special type, e.g.,
that in $N$-qubit cat states.

\section{Discussion and Conclusion}
\label{sec5}

The numerical calculations of the previous section show that the
quantum baker's maps are, in general, good at creating multipartite
entanglement amongst the qubits.  It was found, however, that some
quantum baker's maps can, on average, entangle better than others,
and that all quantum baker's maps fall somewhat short of generating
the levels of entanglement expected in random states.  This might be
related to the fact that spatial symmetries in the baker's map allow
deviations from the predictions of random matrix theory
\cite{balazs}. Such deviations are apparent in the statistics of the
eigenvectors of $\hat{B}_{N,n}$ and might also taint the randomness
of our quantum-baked states. In this light, an entanglement measure
might, in fact, provide a reasonable test for the randomness of
states produced by a quantum map.

In our case, entanglement amongst the qubits relates directly to
correlations between the coarse and fine scales of classical phase
space.  Although we have only considered entanglement in position,
the Fourier transform provides a means to investigate entanglement in
momentum, and the partial Fourier transform (\ref{partialfourier})
might be used for all intermediate possibilities.  These qubit bases,
while naturally embedded in the construction of the quantum baker's
maps, might also be applied to other maps of the unit square and,
hence, to entanglement production in general for quantized mappings
of the torus.

We should expect high levels of entanglement creation in quantum maps
that are chaotic in their classical limit.  Such maps have a
dynamical behavior that produces correlations between the coarse and
fine scales of phase space. This behavior is described classically in
the form of {\it symbolic dynamics}.  Investigations into
entanglement production, using the above product bases, allow us to
characterize the quantum version of such correlations.  To develop a
complete picture, however, the entangling properties of regular
systems must first be addressed.  The possibility of nonentangling
quantum maps, such as $\hat{B}_{N,N}$, which produce stochasticity in
their classical limit, must also be addressed.

We can apply our results to a preliminary investigation of the
relation between entanglement production and the classical limit.  As
remarked previously, sequences of quantum baker's maps for which the
number of momentum bits, $N-n$, is held constant do not approach the
classical baker's map in the limit $N\rightarrow\infty$, but instead
give rise to stochastic variants.  To relate this behavior to
entanglement production, consider the time average of the
Meyer-Wallach entanglement measure $Q$, say,
\begin{equation}
\overline{\langle Q\rangle}\equiv \lim_{m\rightarrow\infty}{1\over m}
\sum_{k=1}^m\langle Q\rangle_k
\end{equation}
where, as before, the average $\langle\:\cdot\:\rangle$ is taken over
a uniform distribution of initial product states.  This quantity
provides the long time saturation value of $Q$ and is plotted for all
quantum baker's maps up to total of $N=10$ qubits in Fig.~\ref{fig11}.
Note that although one observes a drop in the levels of entanglement
for a small number of position bits, $n$, there is no apparent
connection between the level of $\overline{\langle Q\rangle}$ and the
advent of spurious stochastic limits when the number of momentum bits, $N-n$,
remains constant as
$N$ increases.  One might have expected the saturation value to be
suppressed in such limits, but this appears not to be the case.  We
tentatively conclude that entanglement production in the qubit bases
is unrelated to the creation of a stochasticity in the classical
limit.  A similar picture emerges when the subsystem entropy is used
as the entanglement measure.

In conclusion, we have found that the quantum baker's maps are, in
general, good at generating multipartite entanglement amongst qubits.
Given the relation between our qubit bases and classical phase space,
this behavior should be expected to arise whenever such a quantum map
is chaotic in its classical limit.

\begin{acknowledgments}
This work was supported by Office of Naval Research Grant
No.~N00014-00-1-0578.
\end{acknowledgments}

\end{document}